\documentclass[twocolumn]{aastex62}

\definecolor{orange}{rgb}{0.7,0.4,0}

\accepted{in ApJ}

\usepackage{amsmath,amssymb,amstext}
\usepackage{CJK}

\newcommand{\ppm}{\ensuremath{\boldsymbol{p_m}  }}
\newcommand{\fe}{\ensuremath{\mathrm{[Fe/H]}}}
\newcommand{\dfe}{\ensuremath{\nabla\mathrm{[Fe/H]}}}
\newcommand{\tlookback}{\ensuremath{t_\mathrm{lkback}}}

\newcommand{\tobs}{\ensuremath{\tau_\mathrm{obs}}}

\newcommand{\Ro}{\ensuremath{R_0}}

\newcommand{\arexp}{\ensuremath{\alpha_{\mathrm{R_{exp}}}}}
\newcommand{\tsfr}{\ensuremath{\tau_{\mathrm{SFR}}}}
\newcommand{\srm}{\ensuremath{\sigma_\mathrm{RM7}}}
\newcommand{\rfeh}{\ensuremath{R_{\fe =0}^{\mathrm{now}}}}

\newcommand{\gfe}{\ensuremath{{\gamma_\mathrm{[Fe/H]}}}}
\newcommand{\rold}{\ensuremath{R_\mathrm{old}}}			

\newcommand{\pRgRot}{\ensuremath{p(R~|~\Ro , \tau, \ppm) }}

\newcommand{\Gyr}{\,\mathrm{Gyr}}

\definecolor{darkgreen}{rgb}{0.09, 0.45, 0.27}

\newcommand{\rom}{radial orbit migration}

\shorttitle{Inside-out Growth of the MW disk}
\shortauthors{Frankel et al.}

\begin{document}
\begin{CJK*}{UTF8}{gbsn}


\author[0000-0002-6411-8695	]{Neige Frankel}\email{frankel@mpia.de}
\affiliation{Max Planck Institute for Astronomy, K\"onigstuhl 17, D-69117 Heidelberg, Germany}

\author[0000-0003-4593-6788]{Jason Sanders}
\affiliation{Institute of Astronomy, University of Cambridge, Madingley Road, Cambridge CB3 0HA}

\author[0000-0003-4996-9069]{Hans-Walter Rix}
\affiliation{Max Planck Institute for Astronomy, K\"onigstuhl 17, D-69117 Heidelberg, Germany}

\author[0000-0001-5082-9536]{Yuan-Sen Ting(丁源森)}
\altaffiliation{Hubble fellow}
\affiliation{Institute for Advanced Study, Princeton, NJ 08540, USA}
\affiliation{Department of Astrophysical Sciences, Princeton University, Princeton, NJ 08544, USA}
\affiliation{Observatories of the Carnegie Institution of Washington, 813 Santa Barbara Street, Pasadena, CA 91101, USA}

\author{Melissa Ness}
\affiliation{Department of Astronomy, Columbia University, 550 W 120th St, New York, NY 10027, USA}
\affiliation{Flatiron Institute, 162 5th Avenue, New York, NY 10010, USA}

\title{The Inside-out Growth of the Galactic Disk}

%
%
%
%
%
%

\begin{abstract}

We quantify the inside-out growth of the Milky Way's low-$\alpha$ stellar disk, modelling the ages, metallicities and Galactocentric radii of APOGEE red clump stars with $6 < R < 13$ kpc. 
The current stellar distribution differs significantly from that expected from the star formation history due to the redistribution of stars through radial orbit mixing.
We propose and fit a global model for the Milky Way disk, specified by an inside-out star formation history, radial orbit mixing, and an empirical, parametric model for its chemical evolution. We account for the spatially complex survey selection function, and find that the model fits all data well.
We find distinct inside-out growth of the Milky Way disk; the best fit model implies that the half-mass radius of the Milky Way disk has grown by 43\% over the last 7~Gyr. Yet, such inside-out growth still results in 
present-day age gradient weaker than $0.1 \mathrm{~Gyr~kpc^{-1}}$. Our model predicts the half-mass and half-light sizes of the Galactic disk at earlier epochs, which can be compared to the observed redshift
-size relations of disk galaxies.
We show 
that \rom~can reconcile the distinct disk-size evolution with redshift, also expected from cosmological simulations, with the modest present-day age gradients seen in the Milky Way and other galaxies.

\end{abstract}


%
%
%
%
%
%

\section{Introduction}
\label{sec:introduction}

The star formation history is a key formation and evolution aspect for any disk galaxy, as it governs its resulting stellar structure. Subsequent orbit mixing processes can redistribute the stars and remove dynamical memory of their birth conditions. Constraining galaxy evolution requires knowledge of both the initial formation distribution and the importance of orbit mixing.

Galaxy disks are thought to grow from inside-out. \cite{peebles_1969} first postulated that gas disks acquire their angular momenta from tidal torques; with the low angular momentum gas cooling, settling and forming stars on shorter time-scales than the high angular momentum gas \citep{larson_1976}.
(semi-)Analytic hierarchical models for disk formation in a cosmological context \citep[e.g.,][]{fall_efstathiou_1980,mo_mao_white_1998, somerville_etal_2008, dutton_2011} have predicted that galaxy disks must grow from inside-out to reproduce the observed size-luminosity-velocity relations \citep{tully_fisher_1977, courteau_2007}.
Similarly, cosmological and zoom-in simulations show disks with star formation time-scales that increase with the distance to the galactic center or gas accretion of higher angular momentum at later times \citep{avila-reese_etal_2018, grand_etal_2017, aumer_white_naab_2014, brook_etal_2006, brook_etal_2012, pilkington_etal_2012, agertz_2011}.
Measuring the rate at which stellar disks grow can give insight to constrain the interplay between the physical processes involved in gas accretion, cooling and forming stars, as well as the global evolution of disk galaxies.
   
Observationally, a number of independent measurements suggests that disk galaxies grow from inside-out. Several studies indicate that disk galaxies of a given stellar mass are smaller at higher redshift \citep{ferguson_2004, barden_etal_2005, franx_etal_2008, buitrago_2008,van_dokkum_etal_2013, van_der_wel_etal_2014, rodriguez-puebla_etal_2017}. 
Likewise, in massive disk galaxies of the local Universe, integrated light from older stellar populations, as traced by color, is more centrally concentrated than that of younger populations, and star formation is more spatially extended than the overall stellar distribution \citep[e.g.][]{macArthur_2004, munos-mateos_etal_2007,boissier_2008, munoz-mateos_etal_2011, wang_Galex_2011, pezzulli_2015}. 
Similarly, resolved stellar observations of Local Group galaxies (e.g, NGC300, M33 and more recently NGC7793) using the \textit{Hubble Space Telescope} showed that old populations are often more centrally concentrated than young populations \citep{gogarten_etal_2010, williams_etal_2009a, sacchi_2019}. However, this does not appear to be the case in M31 \citep{bernard_etal_2015}. 
Finally, integral field spectroscopy surveys, such as MaNGA \citep{bundy_2015_manga} and CALIFA\citep{sanchez_etal_CALIFA_2012} have led to the detection of mostly weak age radial gradients in massive disk galaxies \citep{goddard_2017, garcia-benito_etal_2017,gonzalezdelgade_etal_2014, perez_2013}.

Most of the described galaxy observations are consistent with inside-out growth, but are restricted to studying present-day galactocentric radii of stellar populations, instead of considering their (unknown) birth sites. Dynamical processes in disk galaxies rearrange the stars, often weakening or erasing any formation gradients \citep[e.g.][]{minchev_2014_mcm2}. Therefore, to state anything quantitative about inside-out growth from these observations requires knowledge of the strength of these dynamical processes. The importance of dynamical heating from non-axisymmetric perturbations such as spiral arms and molecular clouds has long been recognised \citep[e.g.][]{Lacey1984,CarlbergSellwood1985,sellwood2014,Aumer2016}. Transient spiral patterns produce changes in angular momentum and heating around the Lindblad resonances, restructuring the disc \citep{LBK}. \cite{sellwood_binney_2002} recognized that at corotation, a star could be scattered in angular momentum without associated increase in random motion. This process is known as `radial migration' or `churning' and is generally distinguished from `blurring' which describes changes in angular momentum with associated heating. 
A truly radial-migrated population will appear dynamically identical to a population that has not experienced any dynamical mixing \citep{sellwood2014}, making their separation hard. In this work, we ignore kinematics of the stars so cannot distinguish between the different dynamical processes that scatter stars radially, and consider the combined effect as `radial orbit migration'. If radial mixing is strong, then the age gradients expected from inside-out growth can be severely weakened. An old star at a large Galactocentric radius can have formed at a smaller radius, erasing evidence for inside-out growth. All these aspects (inside-out star formation, radial migration) must be accounted for simultaneously in any modelling as demonstrated in \cite{schoenrich_mcmillan_2017, minchev_etal_2019, frankel_etal_2018}.

For the Milky Way, we have an opportunity to disentangle the effect of inside-out formation from 
\rom~as we have access to 
positions, chemical compositions and ages of individual stars from spectroscopic observations. 
Although dynamical processes cause stars to lose dynamical memory of their birth sites, they are thought to retain chemical memory. With assumptions on the past history of the Galaxy, these observations allow the linking of stars to their birth locations from their chemistry or ages through `weak chemical tagging', (different from the classical `chemical tagging' as described in \cite{freeman_bland-hawthorn_2002} and \cite{ting_etal_2015a} for example) such that the dynamical processes can be `rewound' and the formation properties measured \citep[e.g.,][]{schonrich_binney_2009a,sanders_binney_2015,hayden_2015,schonrich_mcmillan_2017,frankel_etal_2018,minchev_2018}.

The recent advance of wide spectroscopic surveys of the Milky Way disk has produced pioneering work in Galactic archaeology enabled by unprecedented samples samples of $\sim10^5-10^6$ stars well beyond the solar neighbourhood. Earlier work considered the distribution of abundances at each Galactic location $p([\alpha/\mathrm{Fe}], \fe ~|~R)$ \citep[e.g.][]{hayden_2015} or $p(R~|~[\alpha/\mathrm{Fe}], \fe)$ \citep[e.g.][]{bovy_etal_2012},  finding that stars with high [$\alpha$/Fe] were more centrally concentrated. But [$\alpha$/Fe] was used as a chemical clock proxy for stellar age, and the focus was on the basic differences between the high-$\alpha$ and the low-$\alpha$ disks.
More recently, stellar ages have become available for many stars, making it feasible to study the age distribution at different locations, and the spatial distributions of mono age populations $p(R~|~ \tau, \fe)$ -- a more explicit measure of the history of the Galaxy \citep[e.g.][]{bergemann_2014, bensby_2010, mackereth_etal_2017}.
However, as the Galaxy evolves it is non-trivial to relate these distributions to the star formation history of the disk. Here, we set out to build a model for where and when stars were born over a large range of Galactocentric radii $p(\Ro. \tau)$, with direct use of stellar ages, and accounting for radial orbit migration. We focus on the low-$\alpha$ disk only (the last ~8 Gyr of the Milky Way evolution), propose an evolution scenario through parametrized equations and fit the parameters using APOGEE data.
 
This paper is the second in a series, developing and applying a framework for a global evolutionary Milky Way disk model introduced in \cite{frankel_etal_2018}, with emphasis on inside-out growth.
We present the data we model in Section~\ref{section_data}. In Section~\ref{section:model}, we describe aspects of the model itself: (1) the survey selection function and (2) the Galactic disk; this model is an extension of \cite{frankel_etal_2018}, where it is described in some detail.  We then present the results of the model fit to the data in Section~\ref{section_results}. Finally, we interpret these in a more global context of galaxy disk formation and evolution and compare them to disk galaxies observed at different redshifts in Section~\ref{section:astrophysical_implications}, and discuss the limitations in Section~\ref{section:limitations}.

%
%
%
%
%
%

%
%
\begin{figure}
    \centering
    \includegraphics[width=\columnwidth]{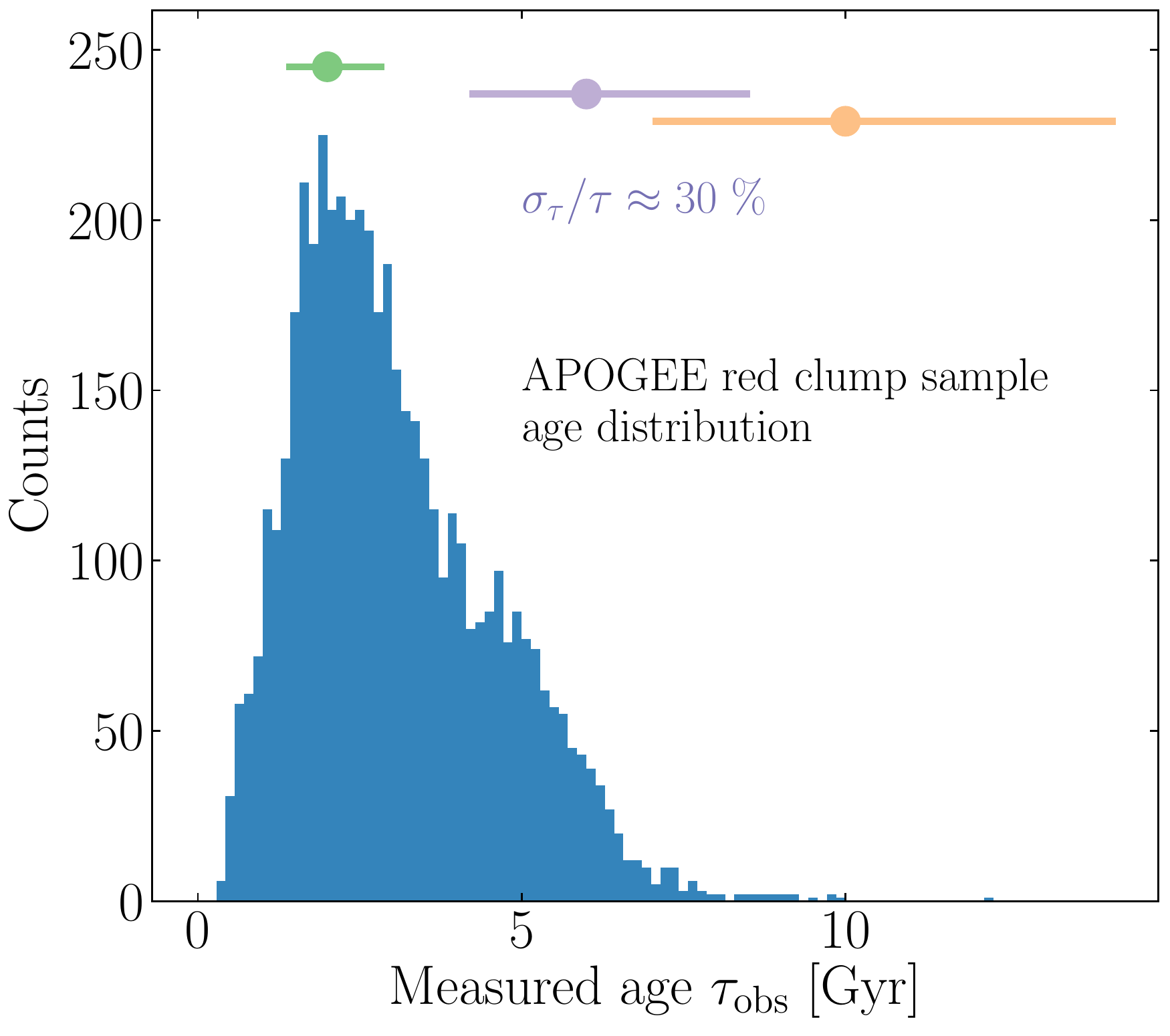}
    \caption{Age distribution of the APOGEE red clump stars from the low-$\alpha$ Galactic disk, used in the analysis presented here. These ages were determined in \cite{ting_rix_2018}, tied to asteroseismic mass estimates. Age uncertainties are $\sigma_{\log_{10} \tau} \approx 0.15$ dex, resulting in about 30\% age uncertainties, as illustrated for three ages: 2 Gyr (green), 6 Gyr (purple), 10 Gyr (orange). The peak in the distribution at 2 Gyr does not reflect the age distribution of the ``underlying'' stellar population, but is expected from theory and can be modelled quantitatively: it reflects the mass dependence -- and thereby age dependence -- of the life time of the core helium burning evolutionary stage that defines red clump stars.}
    \label{fig:age_distribution}
\end{figure}

\begin{figure}
    \centering
    \includegraphics[width=\columnwidth]{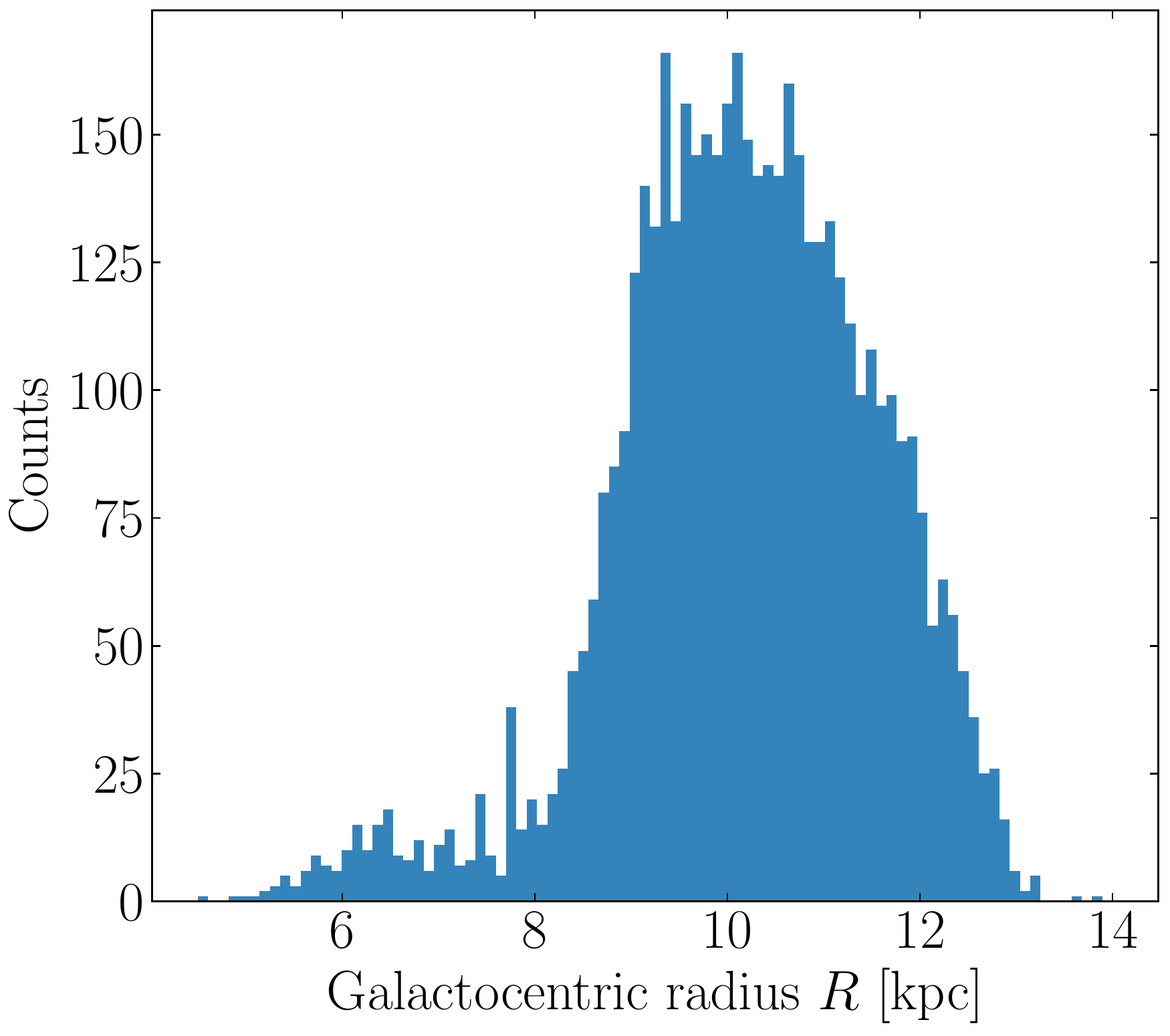}
    \caption{Galactocentric radius distribution of the red clump stars (as in Fig. \ref{fig:age_distribution}), used in this analysis. They span a range of 6~kpc to 13~kpc, with the vast majority of sample stars beyond the Solar radius; this latter property traces back to the Galactocentric radius $R$ distribution of the initial \cite{bovy_etal_2014} parent sample and the subsequent selection cuts we have imposed (Section~\ref{section_data}).}
    \label{fig:R_histogram}
\end{figure}

\section{Data: APOGEE Red Clump Giants}
\label{section_data}

We use asteroseismically calibrated ages, [Fe/H] and 3D positions of red clump stars from the 12th data release of the APOGEE near-infrared spectroscopic survey \citep[Apache Point Observatory for Galactic Evolution Experiment,][]{Alam_2015_apogeedr12, majewski_etal_2017}. Red clump stars are low mass core helium burning stars that have gone through the helium flash. Since they have similar core masses, they have similar luminosities, which make them good standard candles and suitable for Galactic archaeology studies that require precise distances. The core helium burning stage life-time is longer for initially more massive stars. Therefore, the overall red clump population is fairly young, with an age distribution that peaks around 2 Gyr \citep{girardi_2016}, as can be seen in Fig.~\ref{fig:age_distribution}.

\citet{bovy_etal_2014} describes the selection of the red clump population from the APOGEE data. It is based on cuts in stellar parameters $\log g$, $T_\mathrm{eff}$ and infra-red photometry $(J-K)_0$ and $H$. We cross-match this red clump catalog with that of \cite{ting_hawkins_rix_2018} to remove possible contaminants. \cite{ting_hawkins_rix_2018} used a data-driven approach trained on APOKASC2 \citep{pinsonneault_2018} to predict asteroseismic parameters (and hence the evolutionary stage) from stellar spectra, and evaluate their contamination fraction to 3\%.

We further restrict our sample to the low alpha stars as in \cite{frankel_etal_2018}, tracing the most recent evolution of the Milky Way disk, in order to cut down undesired information from older stars. In addition, we consider only the `short cohort' fields of APOGEE, as defined in \cite{zasowski_2013}, which contain the brightest and closest objects (with approximately $7 < H < 12$). Stars further away, in longer `cohorts' because they are fainter, may have higher extinctions, at a level where modelling extinction may be challenging. We restrict our sample to the APOGEE fields where the median extinction of APOGEE stars is less than $\approx 0.6$ in $H$ band. After these cuts, our sample consists of $\sim5381$ stars in 142 APOGEE fields.

We use metallicities $\fe$ known to about $\pm 0.05$ dex and 3D positions ($l, b, D$) with $(l, b)$ standing for Galactic longitude and latitude and with photometric distances $D$, known to about 7\% for such standard candles, from \cite{ness_etal_2016}. The ages $\tau_\mathrm{obs}$ known to about 30\% are taken from \cite{ting_rix_2018}. APOGEE has delivered measurements of 15 chemical abundances, which we could, in principle, use. We choose to restrict to \fe and age, because they are sufficient for our purpose: at given \fe and $\tau$, other abundances [X/Fe] of the low-$\alpha$ disk can be predicted with good precision \citep{ness_etal_2019}. 

The ages were determined using neural networks trained on stars that have asteroseismic age estimates from APOKASC2 \citep{pinsonneault_2018}. The neural networks were trained to predict ages from stellar spectra in \cite{ting_rix_2018}. This age determination method applies the same philosophy as the work described in \cite{ness_etal_2016}, who used a quadratic model to map from stellar spectra to ages rather than non-linear functions. It was shown, in a separate works using respectively asteroseismic ages \citep{Martig2016, silva_aguirre_2018} and Bayesian isochrone fitting \citep{feuillet_etal_2016, feuillet_2018}, that most of the spectral information on the stellar mass (and hence stellar age) comes from surface abundances of the CNO cycle elements brought up during the mass dependent dredge-up process.

Since the largest source of uncertainties comes from the ages (i.e. the age uncertainties are much larger than metallicity and distance uncertainties), we will account for them through a noise model, and treat the distances and metallicities as noise-free variables. 
The observed age distribution of the resulting sample is illustrated in Figure \ref{fig:age_distribution}, and the Galactocentric radius distribution is shown in Figure \ref{fig:R_histogram}. Most of the stars of our sample are young and located in the outer 8-12 kpc of the Milky Way disk. Therefore, our modelling will describe mainly the recent evolution of the outer disk.
Our final data set consists of these 5381 stars with $\mathcal{D} = \{l, b, D, \fe, \tobs\}$.

%
%
%
%
%
%
\section{Modelling the Data Set} \label{section:model}

%
%
\begin{figure}
    \centering
    \includegraphics[width=\columnwidth]{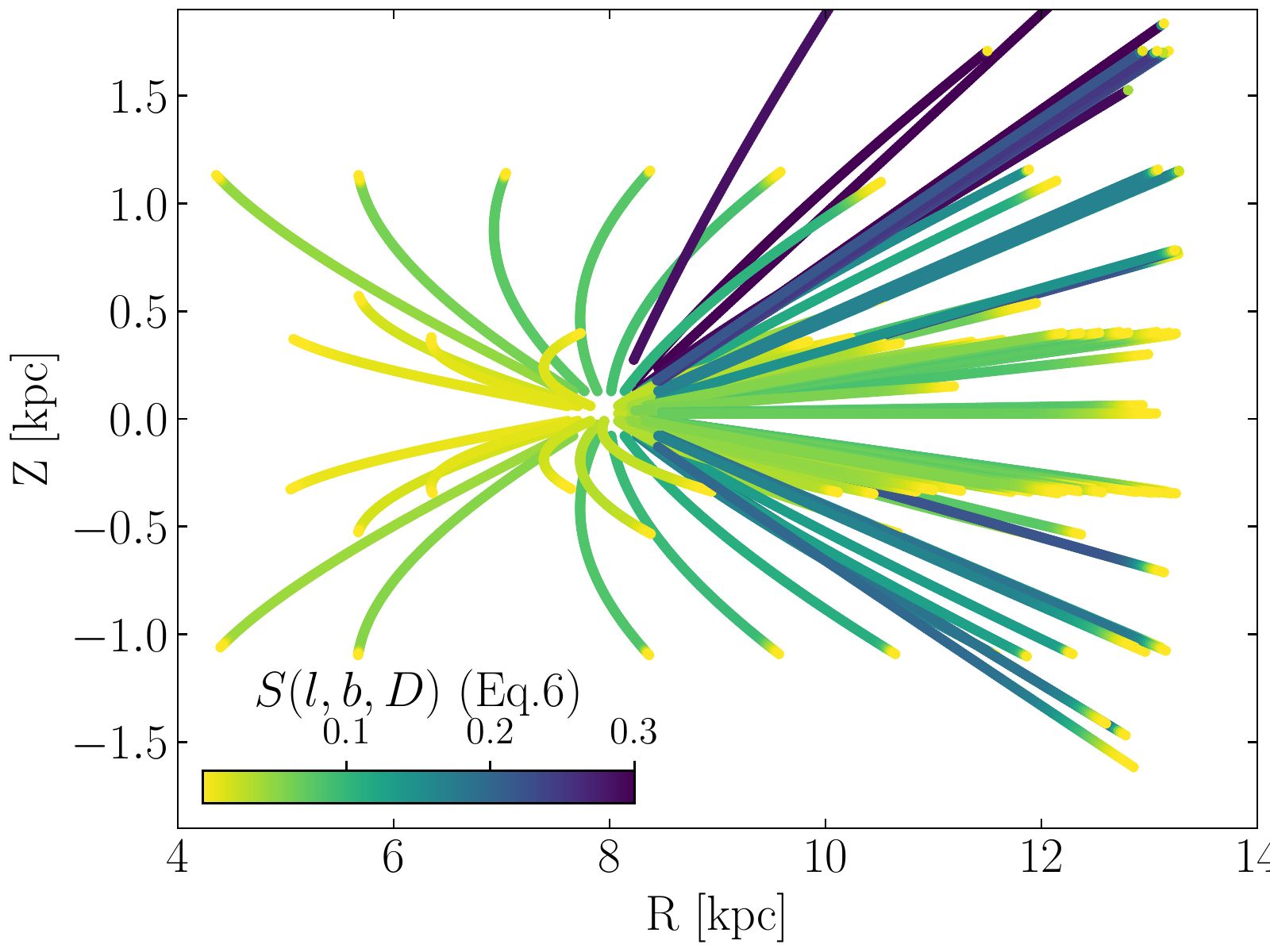}
    \caption{Distance-dependent fraction of observed red clump stars in the APOGEE pointings used here. 
    This is the product of the fraction of photometrically eligible stars in each of the 142 fields that were spectroscopically targetted (Eq.~\ref{eq:sel_frac}), and the distance-dependent probability that a star of red clump luminosity does not have prohibitive dust extinction, as calculated from the \cite{green_etal_2018} extinction map (Eq.~\ref{eq:selfrac_dust}).}
    \label{fig:selection_function}
\end{figure}

\begin{table*}
    \centering
    \begin{tabular}{lll}
         \hline
         Symbol & Name & Type and appearance in the text\\
         \hline
         $l, b, D$ & Galactic longitude, latitude, distance from the Sun            & Assumed noise-free observables \\
         $X, Y$ & Cartesian coordinates for stellar positions in the plane          & Disk model variables \\
         $R$, $Z$ & Galactocentric radius, height above the Galactic plane          & Disk model variables \\
         \Ro & Galactocentric radius of a star at its birth                         & Disk model variable\\
         \fe & Metallicity of stars                                                 & Assumed noise-free observable\\
         \tobs & Measured stellar age                                               & Observable that has errors \\
         $\tau$ & Modelled true age                                                 & Disk model variable \\
         \hline
         $\sigma_{\log_{10}\tau}$ & Uncertainties in measured log age   & Noise model parameter, fixed \\
         $R_d$ & Integrated scale-length of the star-forming disk   & Element of \ppm~ in Eq.~\ref{eq:surface_birth_profile}, to fit\\
         \tsfr, $\tau_m$, $x$ & Star formation time-scale, onset time, inside-out degree & Elements of \ppm~in Eq.~\ref{eq:SFH_def}, to fit \\
         $h_z(a_z)$ & Disk scale-height as a function of the scaling parameter $a_z$ & Nuisance parameter in Eq.~\ref{eq:scale-height}, to fit\\
         $R_\mathrm{old}$ & Scale-length of the disk for $\tau > \tau_m$ & Nuisance parameter in Section~\ref{subsection:old_component}, to fit \\
         \srm & Strength of radial orbit mixing & Element of \ppm~in Section \ref{subsection:radial_migration}, to fit \\
         \gfe & chemical evolution parameter time dependency & Nuisance parameter in Eq.~\ref{eq:enrichment_time}, to fit\\
         \rfeh & Radius of solar metallicity in the star-forming gas & Nuisance parameter in Eq.~\ref{eq:eq_Fe_Ro_t}, to fit\\
         \dfe & [Fe/H] radial gradient in the star-forming gas & Nuisance parameter in Eq.~\ref{eq:eq_Fe_Ro_t}, to fit\\
         \hline
         $\Omega_i$ & Solid angle of APOGEE pointing i & Selection function parameter, fixed\\
         $H_\mathrm{RC}$ & Red Clump absolute magnitude in H band & Selection function parameter, fixed\\
         $A_H$, $A_K$  & Extinction in H and K bands & Selection function variable\\
         $(J-K)_0$ & Deredenned color of a star & Selection function variable  \\
         \hline
         \end{tabular}
    \caption{Main variables and parameters used in the model.}
    \label{table:variables_parameters}
\end{table*}

We set out a global model for the data described above: $\mathcal{D} = \{l, b, D, \fe, \tobs \}$ and their uncertainties. We make a clear difference between the measured age $\tobs$ and the true age $\tau$ that we use for the modelling, to account rigorously for age uncertainties (see Section \ref{subsection:noise_model}). For clarity, all the model variables and parameters are summarized in Table \ref{table:variables_parameters}. We start by splitting this model in three main terms: (1) the model for the selection of the disk stars in the APOGEE survey, (2) a global forward model for the Galactic disk with model parameters in an array \ppm~and (3) the normalization constant over the observable space limited by the selection as in \cite{rix_bovy_2013}.
\begin{equation}\label{eq:model_dataset}
p(\mathcal{D}~|~ \ppm, ~\mathrm{selection})  =  \frac{S(l,b,D)f_\mathrm{RC}(\tau) p(\mathcal{D} ~|~\ppm)}{V_s(\ppm)}
\end{equation} 
where $p(\mathcal{D} ~|~\ppm, ~\mathrm{selection})$ is the normalized model of the data set, $V_S(\ppm)$ is a normalization constant, $p(\mathcal{D} ~|~\ppm)$ is the physical model for the Galactic disk, and $S(l,b,D) = p(\mathrm{selection}~|~l, b, D)$ is the survey selection function: the probability that a star ends up in the catalog, given its properties: position $(l, b)$, magnitude $H$ or distance $D$ in our particular case for a standard candle, deredenned color $(J-K_s)_0$, and integrated extinction $A_H(D)$ along the line of sight (Section \ref{subsection:selection_function}). The term $f_\mathrm{RC}(\tau)$ is the fractional stellar mass contained in the red clump population at given age. It can be determined assuming an initial mass function and using stellar isochrones. Generally, selecting stellar populations through cuts in observables implies biasing the sample in age and metallicity. For the red clump population, \cite{bovy_etal_2014} showed that $f_\mathrm{RC}$ is as strong function of age, and a weak function of metallicity. Therefore, we neglect the weak metallicity dependence and adopt the fit of $f_\mathrm{RC}$ as a function of age from \cite{bovy_etal_2014}.

The normalization constant (or survey volume) can be computed by integrating the unnormalized model over the observable space:
\begin{equation} \label{eq:survey_volume_def}
V_S(\ppm) =  \int _\mathcal{D} p(\mathcal{D} ~|~\ppm) S(l,b,D)f_\mathrm{RC}(\tau) \mathrm{d}\mathcal{D},
\end{equation}
which is a 5 dimensional integral over all the physical properties of the data. The integrals over $(l, b)$ are obtainable analytically and can be transformed into a sum over APOGEE fields, assuming that the properties of stars do not vary in $(l, b)$ over a single APOGEE pointing (of about 1.5 degree radius). We evaluate this integral in subsection \ref{subsubsection:survey_volume}. Expanding out the data and writing the spatial distributions in Cartesian coordinates ($X, Y, Z$), we have
\begin{equation}\label{eq:model_jaco}
\begin{split}
p(l,& b, D, \tobs, \fe ~|~\ppm) \\
& =  D^2 \cos(b) p(X, Y, Z, \tobs, \fe ~|~\ppm),  
\end{split}
\end{equation}
where $p(X, Y, Z, \tau, \fe~|~\ppm)$ is our model for the Galactic disk described in Section \ref{subsection:model}, and its relation to the similar term in Eq.~\ref{eq:model_jaco} is the convolution over age uncertainties as described in Section \ref{subsection:noise_model}.



%
%
%
%
\subsection{Modelling APOGEE Selection Function \label{subsection:selection_function}}

The APOGEE survey targeted stars in different lines of sights, or pointings, centered on directions $(l, b)$. Therefore, a spatial histogram of the stars contained in the APOGEE catalog does not reflect the spatial density of stars in the Milky Way disk. Using these stars to infer the spatial structure of the Milky Way disk requires forward modelling of this selection process.

The details of the APOGEE-1 selection function are described in \cite{zasowski_2013}. We apply and summarize here the general method laid-out in \cite{rix_bovy_2013, bovy_etal_2014} and refer the reader to these references for details \footnote{we have additionally made the selection function for fields used here available as a fits table at \url{https://github.com/NeigeF/apogee_selection_function} with a tutorial}. For most stars of the APOGEE sample, we can assign a probability that this star was observed given its location, magnitude and color (selection function), which is then combined with a probability for this star to be at this location (the density model). In most APOGEE disk fields, stars were selected according to simple magnitude cuts in the $H$ band and cuts in dereddened $(J-K)_0$ color from the photometric sample 2MASS \citep{skrutskie_etal_2006} which is assumed complete within these cuts.
The de-reddened colors were obtained in \cite{zasowski_2013} using the Rayleigh Jeans Color Excess method \citep{majewski_RJCE_2011}, based on combinations of photometry in different near- and mid-infrared bands and on the assumption that most stars have a similar intrinsic color in the Rayleigh-Jeans part of their spectrum.
When too many stars in a pointing satisfy these criteria, the subset of stars to observe was drawn roughly randomly from the 2MASS. This subset contains a fraction $S_\mathrm{i}$ of all available stars, $\mathrm{\star}$,
\begin{equation}\label{eq:sel_frac}
    S_\mathrm{i} = \frac{\mathrm{\#\star ~in ~APOGEE ~field ~i}}{\mathrm{\# \star ~in ~2MASS ~matching ~selec. ~criteria ~in ~i}}.
\end{equation}
We focus on the main disk fields of APOGEE that are not dominated by ancillary programs (Eq.~\ref{eq:sel_frac} is approximately valid only for targets drawn randomly from 2MASS, not for targets chosen for a specific purpose). We also remove the fields for which the selection criteria were complex to model (see a detailed description of the numerous particular cases in \cite{zasowski_2013}) or irrelevant to the scope of the present work, e.g., halo fields which contain fewer disk stars. 

In addition, a fraction of stars in each field in the Milky Way disk is too extinguished by interstellar dust to be detectable within the magnitude limits of APOGEE \citep{bovy_rix_etal_2016}. This is a function of distance, and can be modelled, if we have a model for the extinction spatial distribution in $H$ band. We use the Bayestar17 3D extinction map \citep{green_etal_2018}, which predicts the extinction distribution in different bands at any $(l, b, D)$. We convert the map's output values to $H$ band using the extinction coefficient $0.468$ and assume the extinction law of \cite{indebetouw_2005} ($A_H/A_K=1.55$) to convert the K band extinction values stated by APOGEE, to the H band. Assuming red clump stars are standard candles of magnitude $H_\mathrm{RC} = -1.49$ \citep{laney_2012}, the probability that a red clump star can be seen at a given distance D within APOGEE magnitude limits $H_\mathrm{min}$ and $H_\mathrm{max}$ is the fractional area of the field at this given distance that is not hidden by dust extinction $A_H$.
The map's highest resolution (3.4') allows computation of this ratio by counting the distance-dependent fraction of pixels that satisfy the inequality
\begin{equation}\label{eq:selfrac_dust}
    f_i(D ~|~ \mathrm{field ~i}) = \frac{\mathrm{\# ~ pix}( H_\mathrm{min} < H(D, H_\mathrm{RC}, A_H) < H_\mathrm{max} )}{\mathrm{\# ~pix ~in ~ field~i }}.
\end{equation}
We can finally recast this into the overall selection function, dubbed `\textit{effective selection function}' when it accounts for both the probability to select a star from the photometric sample and for the probability to see a star given dust extinction, plotted in Figure \ref{fig:selection_function},
\begin{equation}\label{eq:selection_function_dust}
    S(l_i, b_i, D) = S_i \times f_i(D  ~|~ \mathrm{field ~i}).
\end{equation}
This is one of the three terms to cast in Eq.~\ref{eq:model_dataset}, and we are left with the two others: global model for the Galactic disk, and normalizing integral.

\begin{figure*}
    \centering
    \includegraphics[scale=0.505]{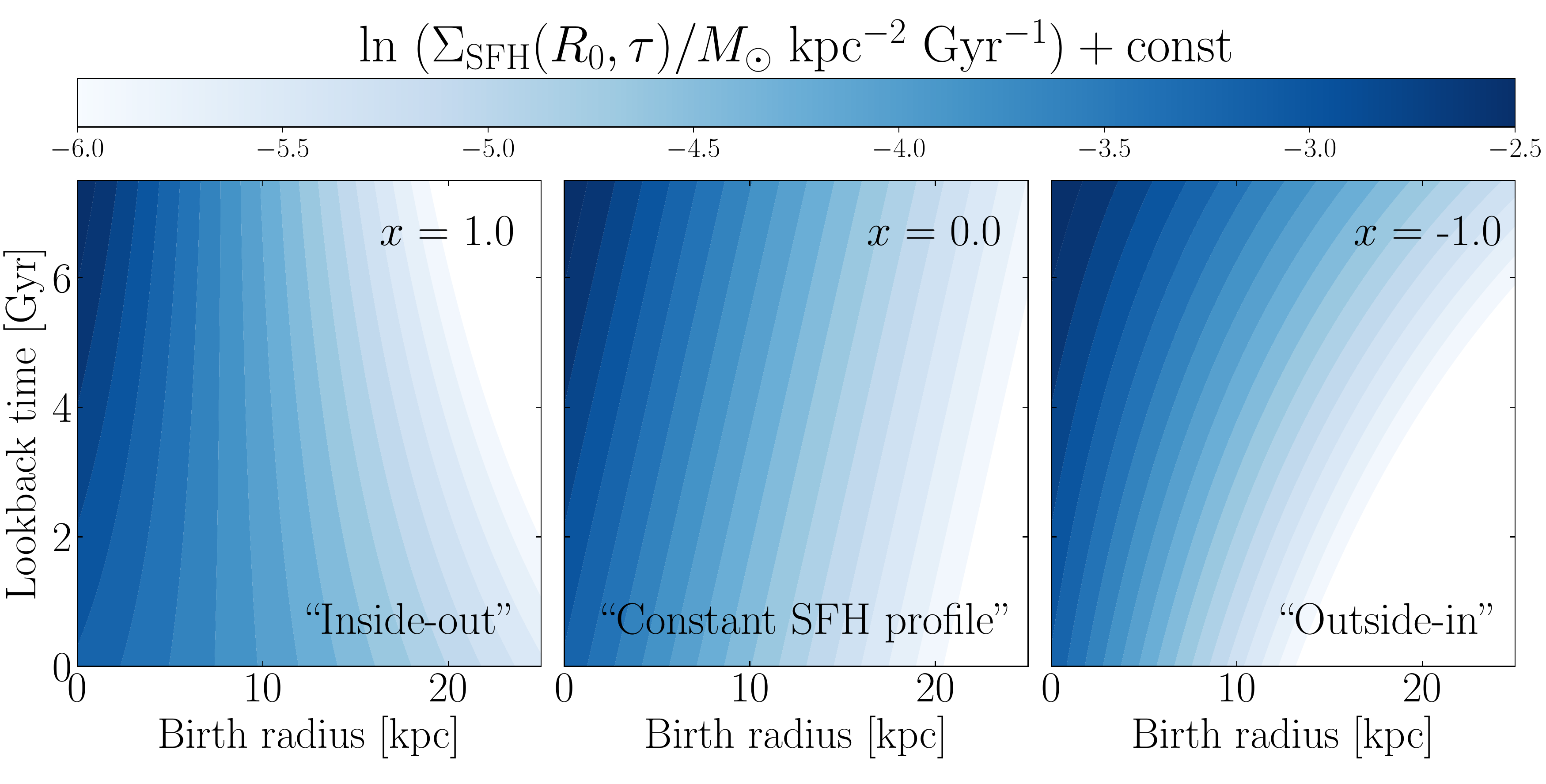}
    \caption{Model star formation rate as a function of time and birth Galactocentric radius for different scenarios considered here. We show the star formation rate surface density $\Sigma_\mathrm{SFR}(\Ro, \tau ~|~ \tsfr, \tau_{m}, x)$, normalized to yield unit total stellar mass, as a function of birth radius $\Ro$, lookback time $\tau$ and inside-out growth degree $x$ as defined in Eq.~\ref{eq:SFH_def}. The panels reflect star formation histories for different `inside-out growth parameters' $x$, from left to right: strong inside-out growth ($x = 1$) that would predict a
    flat star formation history at 8 kpc,
    uniform star formation history ($x = 0$),  and strong outside-in growth, where the disk formed stars initially on a larger scale-length than at present.}
    \label{fig:model_SFH}
\end{figure*}

\subsection{Global Model for the Galactic Disk Evolution\label{subsection:model}}

We aim to model the Galactic disk in terms of the distributions $p(\tau, \fe, R, Z ~|~\ppm)$ of ages, metallicities, Galactocentric radii and heights above the plane. To fit the large data set efficiently, we build the different model aspects (star formation history, enrichment, radial mixing) from parametrized functional families. These are not designed to describe the exact physics in a self-consistent manner, but are physically plausible, fast to compute and have parameters that can be physically interpreted. This methodology permits us to disentangle different effects at play and measure the scales on which they happen. Applying the probabilistic chain rule and marginalizing over the birth Galactocentric radius \Ro, we introduce the different model aspects:
\begin{equation}
\label{eq_p_model}
\begin{split}
p(\tau &, \fe, X, Y, Z ~|~\ppm) \\ 
& = \int p(\Ro, \tau ~|~ \ppm)p(\fe~|~\Ro, \tau,\ppm) \\ & \times p(X, Y, Z ~|~\Ro, \tau,\ppm) \mathrm{d}\Ro.
\end{split}
\end{equation}
The first term in the integral is the birth radius-age distribution of stars and is closely related to the star formation history, (subsection \ref{subsection:SFH}). The second term is the distributions of metallicity in the star forming gas as a function of time, which we adopt from \cite{frankel_etal_2018} and re-fit. The last term is the spatial density resulting from radial orbit migration and vertical heating. It can be split into a radial component (the main focus here) and a vertical component, which depends on the distance from the Galactic mid-plane $Z$:
\begin{equation}\label{eq:pXYZgR0t}
\begin{split}
p(X, Y, Z ~|~\Ro, \tau,\ppm) &= p(X, Y ~|~\Ro, \tau,\ppm) \\ & \times p(Z ~|~\Ro, R, \tau,\ppm).
\end{split}
\end{equation}
The first term on the right hand side is the present-day surface density of stars of true age $\tau$ and birth radius \Ro~ (subsection \ref{subsection:radial_migration}), and the second term is the present-day vertical profile of the disk resulting from the vertical heating of stellar orbits between their birth radius and their present day radius over time (subsection \ref{subsection:vertical_distribution}).\newline


\subsubsection{Radial Dependent Star Formation History \label{subsection:SFH}}

We model the distribution of stars at birth as
\begin{equation}
    \Sigma_\mathrm{SFR}(\Ro, \tau ~|~ \ppm) = \Sigma(\Ro ~|~\ppm)\mathrm{SFH}(\tau ~|~ \Ro, \ppm),
\end{equation}
where the time integrated surface density profile at birth is an exponential in birth radius \Ro
\begin{equation}\label{eq:surface_birth_profile}
\Sigma(\Ro ~|~ \ppm) \propto \exp{\left(-\frac{\Ro}{R_d}\right)},
\end{equation}
and the radially-dependent normalized star formation history SFH is a function of age $\tau$
\begin{equation}\label{eq:SFH_def}
    \begin{split}
            \mathrm{SFH}(\tau ~&|~\Ro, \ppm) = c(\Ro, \ppm) \\ 
            &\times \exp\left[\frac{1}{\tsfr}\left((1-x\frac{\Ro}{\mathrm{8 kpc}})\tau - \tau_m \right) \right].
    \end{split}
\end{equation}
The inside-out growth is encoded in the dimensionless parameter $x$. If $x = 0$, the star formation time-scale is constant across the disk: no inside-out growth. If $x > 0$, stars form on a shorter time-scale in the inner disk than in the outer disk: there is inside-out growth. If $ x < 0$, the star formation time-scale decreases to the outer disk: this would be outside-in growth. We fit for $x$ by maximizing the likelihood, without imposing any prior. The values taken by $\Sigma_\mathrm{SFR}(\Ro, \tau ~|~ \ppm)$ for three different values of $x$ (1, 0 and -1) are illustrated in Figure \ref{fig:model_SFH}. 
More generally, $x$ sets the strength of the linear dependency of the star formation history with Galactocentric radius. For positive $x$ and \tsfr, there is always a Galactocentric radius $R_\mathrm{SFR0}$ such that the star formation rate is a decreasing function of time inside $R_\mathrm{SFR0}$, and is an increasing function of time outside $R_\mathrm{SFR0}$. In the particular case where $x=1$, the star formation history is constant at 8 kpc, decreasing in the inner 8 kpc, and increasing in the outer 8 kpc. This is illustrated in the left panel of Figure \ref{fig:model_SFH}, where at the radius $\Ro = 8$ kpc, the contour lines are vertical.
Finally, the normalization constant $c(\Ro, \ppm)$ is such that the SFH in Eq.~\ref{eq:SFH_def} is normalized between 0 and $\tau_m$.
The model parameters to fit for are \{$\tau_\mathrm{m}$, \tsfr, $x$, $R_d$\}.
This equation is related to the first term of Eq.~\ref{eq_p_model} through
\begin{equation} \label{eq:pR0t}
    p(\Ro, \tau ~|~ \ppm) = 2 \pi \Ro~ \Sigma_\mathrm{SFR}(\Ro, \tau ~|~ \ppm).
\end{equation}

\subsubsection{Radial Orbit Migration \label{subsection:radial_migration}}

Since the present-day Galactocentric radii of stars may be different from their birth positions, we model the effect of radial orbit migration in order to `rewind' back to the birth properties of stars.
We use the radial orbit migration prescription from \cite{frankel_etal_2018} for the probability of a star moving from birth radius \Ro\ to current radius $R$ in a time $\tau$. This is modelled as a global diffusion process with a diffusion strength $\sigma (\tau) = \srm \sqrt{\tau/\mathrm{7~Gyr}}$,
\begin{equation} \label{eq:eq_radial_migration}
\pRgRot = N(\ppm, \tau, \Ro) \exp{\Bigl (-\frac{(R-\Ro)^2}{2 ~\sigma ^2(\tau)}\Bigr )}
\end{equation}
with $N(\ppm, \tau, \Ro)$ a normalization constant. We re-fit for \srm.

The surface density profile of migrated stars of age $\tau$ and from birth radius \Ro~is
\begin{equation}
    p(X, Y~|~ \Ro, \tau, \ppm) = \frac{1}{2 \pi R}p(R~ |~ \Ro, \tau, \ppm),
\end{equation}
and is the first term in Eq.~\ref{eq:pXYZgR0t}.

\subsubsection{Vertical Distribution of Stars\label{subsection:vertical_distribution}}

Since the present work focuses on the radial structure of the Milky Way disk, we are essentially not interested in its vertical structure. However, we must not ignore it as the survey selection function is three dimensional. We therefore adopt a description for the vertical profile that is good enough to characterize the vertical heating history of the disk, and for which we only fit a single parameter.

The Milky Way disk scale-height depends on the ages of stellar populations, and the vertical profile of populations of given age flares \citep[e.g.,][]{mackereth_etal_2017, bovy_rix_etal_2016}. We model the vertical structure of the Milky Way disk as a function of age and Galactocentric radius, using the approximation of an isothermal disk and the harmonic limit \citep[e.g.,][]{BT08}:
\begin{equation}
\begin{split}
p(Z~&|~R, \Ro, \tau, \ppm) = \\ &\frac{1}{2h_z(R, \Ro, \tau)}\mathrm{sech}^2\left(\frac{Z}{h_z(R, \Ro, \tau)}\right),
\end{split}
\end{equation}
where the scale-height is modelled as
\begin{equation}\label{eq:scale-height}
h_z(R, \Ro, \tau) = a_z \sqrt{\frac{2 \overline{J_z}(R, \Ro, \tau)}{\nu (R)}}.
\end{equation}
We fit for the dimensionless parameter $a_z$ and adopt the terms in the square root from the literature:  $\overline{J_z}(R, \Ro, \tau)$ is the mean vertical action of stars of age $\tau$ born at $\Ro$ and now at Galactocentric radius $R$. \cite{ting_rix_2018} have studied the vertical heating history of the Galactic disk using an APOGEE red clump data set and Gaia data \citep{gaia_DR2_ref_brown_2018, gaia_dr2_ref_astrosol_Lindegren_2018}, and provided a functional form for $\overline{J_z}(R, \Ro, \tau)$, which we adopt here.
We calculate the vertical frequency $\nu$ such that $\nu ^2 = \frac{\partial^2\Phi}{\partial^2 z}$ using the  MWPotential2014 potential of the Galpy python package \citep{bovy_2014_galpy}, which is the same gravitational potential $\Phi$ used by \cite{ting_rix_2018} to infer the orbital actions of their data. 



\subsubsection{Chemical Evolution}
We model the chemical evolution of the gas as in \cite{frankel_etal_2018}: the metallicity of a star at birth is modelled as a simple function of its birth Galactocentric radius $\Ro$ and time after birth $\tau$:
\begin{equation}\label{eq:eq_Fe_Ro_t}
\begin{split}
\fe = & F_m - (F_m + \dfe \rfeh) f(\tau )   \\ &+ \dfe  R
\end{split}
\end{equation}
with the time dependency
\begin{equation}\label{eq:enrichment_time}
f(\tau ) = \left( 1 - \frac{\tau}{12~\mathrm{Gyr}} \right)^\gfe.
\end{equation}
We fit for the model parameters \{\dfe, \rfeh, \gfe \} and keep $F_m=-1$ dex fixed. This chemical evolution model makes the following assumptions:
\begin{itemize}
\item there is a tight, unique relation between the birth location and time and the metallicity of a star,
\item there is always a negative radial metallicity gradient in the ISM (modelled through \dfe )
\item at any Galactocentric radius, the \fe\ of the ISM only increases with time (modelled through the exponent \gfe).
\end{itemize}
These assumptions are supported by several chemical evolution models for the late evolution of the Milky Way disk (the past 8 Gyr) after the last major merger \citep[e.g.][]{schonrich_binney_2009b, grisoni2018}. Since this model is purely parametric and fitted to the data, it has the important advantage of bypassing the large uncertainties currently present in chemical evolution models, for example the coupling between the star formation history, possible gas inflows, outflows, radial flows, enrichment sources, supernovae progenitors, supernovae models and nucleosynthesis yields. However, the shortcoming of this model is that it does not contain a self-consistent link between the chemical evolution description and the star formation history. Our approach comes closer to `weak chemical tagging', where $\fe(\Ro, \tau)$ is used to tag stars to their possible birth radius \citep[e.g.][]{minchev_2018, schonrich_binney_2009b, sanders_binney_2015}.

\subsubsection{Accounting for an ``Old'' Disk Component\label{subsection:old_component}}

The evolution model presented above may not be valid at early times in the evolution of the Milky Way; yet, the size-able age uncertainties mean that we have to incorporate the existence of an ``old'' (still low-$\alpha$) disk component.  
We aim to introduce an uninformative model for this old low-$\alpha$ disk by presuming there is
an old star fraction $\epsilon$ in our data set with an uniform age distribution between $\tau_m$ and 12 Gyr, where $\tau_m$ remains part of the formal model fit.
After some experimentation, we have adopted empirically $\epsilon =0.05$, as this leads to astrophysically sensible $\tau_m$.

We model the old star metallicity distribution as a Gaussian centered on solar metallicity with 0.2 dex spread (inspired from \cite{frankel_etal_2018}), the radial distribution as an exponential of scale-length $R_\mathrm{old}$ for which we fit, and the vertical distribution as a {\it sech2} function of scale-height $h_\mathrm{old} = 0.85$ kpc (which is roughly consistent with the local scale-height of old stars \citep{mackereth_etal_2017} ). We do not fit for the vertical scale-height of the older stars, but instead use prior knowledge that old stars have generally dynamically hotter orbits (so greater scale-heights) than younger stars. This makes our separation between `young' and `old' better informed in the presence of large age uncertainties at large ages.

\subsection{Model for Age Uncertainties \label{subsection:noise_model}}
To account for age uncertainties, we convolve the evolution model with a noise model: 
\begin{equation}\label{eq:integral_p_obs}
\begin{split}
p(\tobs, &\mathcal{D'} ~|~ \ppm) = \\
&\int_0 ^{\tau_m} p(\tau, \mathcal{D'} ~|~ \ppm)p_\mathrm{obs}(\tobs ~|~\tau, \sigma_{\log_{10}\tau}) \mathrm{d} \tau
\end{split}
\end{equation}
where here $\mathcal{D'}$ is all assumed noise-free observables (position and metallicity, see Table \ref{table:variables_parameters}), and $\sigma_{\log_{10}\tau}$ is our noise parameter. We assume that age uncertainties are Gaussian in log age with a spread of $\sigma_{\log_{10}\tau} = 0.15$ dex, which we calculated by measuring the standard deviation of the difference in inferred log age and APOKASC2 test set. $p_\mathrm{obs}\left(\log_{10}(\tobs) ~|~\tau, \sigma_{\log_{10}\tau}\right)$ represents the probability of measuring an age $\tobs$ given the true age $\tau$:
\begin{equation}\label{eq:noise_model}
    p_\mathrm{obs}\left(\log_{10}(\tobs) ~|~\tau, \sigma_{\log_{10}\tau}\right) \sim \mathcal{N}\left(
    \log_{10}(\tau), \sigma_{\log_{10}\tau}\right).
\end{equation}
The integral in Eq.~\ref{eq:integral_p_obs} must be computed numerically for each of the 5381 stars, which makes the fitting procedure computationally expensive.

\subsection{Normalization of the Probability Density Function: Survey Volume \label{subsubsection:survey_volume}}
The probability density function must be normalized over the observables.
The survey volume, as defined in Equation \ref{eq:survey_volume_def}, is
\begin{equation}\label{eq:survey_volume_integral}
\begin{split}
& V_S(\ppm) = \int_\mathcal{D''} p(\mathcal{D''} ~|~\ppm) S(l,b,D) \mathrm{d}\mathcal{D''}\\
&= \sum _{\mathrm{i}}^\mathrm{fields} \iiint p(\tau, \Ro, X, Y, Z ~|~\ppm) S_i(D) \Omega_\mathrm{i} D^2 \mathrm{d}D  \mathrm{d}\Ro  \mathrm{d}\tau,
\end{split}
\end{equation}
with $\mathcal{D''} = \{l, b, D, \Ro, \tau\}$ and $\Omega_\mathrm{i}= \iint_\mathrm{field~i} \cos(b)\mathrm{d}l\mathrm{d}b$ the solid angle of plate $i$. In the second line of Eq.~\ref{eq:survey_volume_integral}, we sum over fields instead of integrating over the entire sky ($S$ is zero outside the fields), and we have performed the integral over $(l, b)$ in each field assuming the density varies slowly across the (small) angular size of the field. We have also implicitly integrated over metallicity as $\int p(\fe | \Ro, \tau) \mathrm{d}\fe = 1$ and the selection function is assumed independent of metallicity.
We compute the sum of the remaining 3D integrals using trapezoidal integration on a regular grid of $[D, \Ro, \tau]$ with 42, 38 and 36 points in each dimension respectively. $V_S(\ppm)$ is a function of the model parameters only, so is only evaluated once per optimization step and not for each star.

\subsection{Constructing the Likelihood}

We can now cast the model aspects back into Eq.~\ref{eq_p_model}, and build Eq.~\ref{eq:model_data set} from Equations \ref{eq:model_jaco} and \ref{eq:selection_function_dust}. 
In practice, our model is a mixture of the evolution model and the old, uninformative model:
\begin{equation}
\begin{split}
    p_\mathrm{tot}(\mathcal{D}_i ~|~ \ppm, \mathrm{selection}) = &(1 - \epsilon)\cdot p(\mathcal{D}_i ~|~ \ppm, \mathrm{selection}) \\
    &+ \epsilon \cdot p_\mathrm{old}(\mathcal{D}_i ~|~ \ppm, \mathrm{selection}),
\end{split}
\end{equation}
with an `old fraction' $\epsilon = 5\%$.
Assuming all measurements are independent, we construct the log likelihood $\mathcal{L}$ that the model parameters \ppm ~generated the data $\mathcal{D}$ from our model
\begin{equation}
    \mathcal{L}(\ppm, \{\mathcal{D}\}) = \sum_{i} \ln p_\mathrm{tot}(\mathcal{D}_i ~|~ \ppm, \mathrm{selection}).
\end{equation}
To optimize calculations, we compute the survey volume once per optimization step as it is not a function of the observables. We perform a maximum likelihood estimate of the parameters \ppm~of the global model for all the 5381 stars. We are mainly interested in three parameters: the star formation time-scale, the inside-out growth parameter $x$ and the scale-length of stars at birth $R_d$. However, we fit for all ten parameters in the model (summarized in Table \ref{table_best_fit}) because the global model needs a good description of the Milky Way disk in all aspects in order to describe inside-out growth correctly.

%
%
%
%
%
%

\section{Results}\label{section_results}

\begin{figure}
    \centering
    \includegraphics[width=\columnwidth]{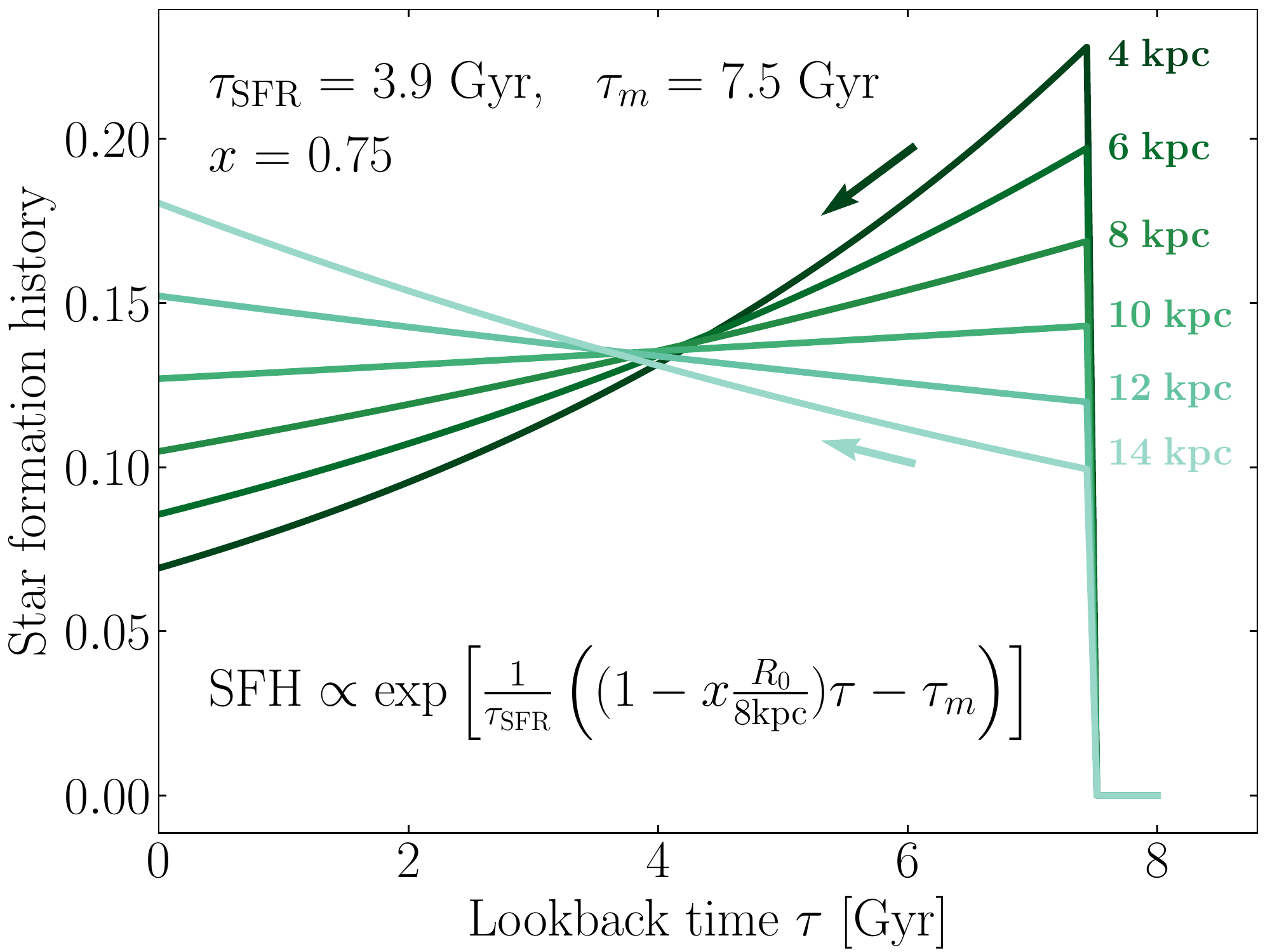}
    \caption{Star formation history $\mathrm{SFH}(\Ro ~|~ \tau)$ implied by the best-fit model at different Galactocentric birth radii $\Ro$ (from dark to light: 4, 6, 8, 10, 12 and 14 kpc). The best fit model has parameters $x = 0.75$, $\tau_m = 7.5\Gyr$ and $\tsfr=3.9$ Gyr. This yields a star formation rate decreasing with time in the inner disk (dark green arrow), flat at 10.5 kpc, and increasing with time in the outer disk (light green arrow).}
    \label{fig:SFH_bestfit}
\end{figure}

The parameter values that maximize the likelihood are presented in Table \ref{table_best_fit}, where the uncertainties quoted are determined from 10 samples of bootstrapped data. We comment below the direct implications and describe the main tests carried to verify the results.

\begin{deluxetable}{lll}
\tablecolumns{3}
\tablecaption{Maximum Likelihood Estimates of \ppm} \label{table_best_fit}
\tablehead{\normalsize Parameter & \normalsize \ppm & 	\normalsize	Best fit}
\startdata
\normalsize Inside-out growth & \normalsize x & \normalsize $0.75\pm 0.17$ (0.69)\tablenotemark{a} \\
\normalsize Star formation time-scale & \normalsize $\frac{\tsfr}{\mathrm{Gyr}}$& 		\normalsize	$3.9\pm 1.5$ (1.4) \\
\normalsize Star formation onset & \normalsize $\frac{\tau_m}{\mathrm{Gyr}}$ & \normalsize $7.5\pm 0.5$ (7.6)\\
\normalsize Disk scale-length & \normalsize $\frac{R_{d}}{\mathrm{kpc}}$&    \normalsize  $3.1\pm 0.4$ (2.9) \\
\normalsize Radial orbit migration &\normalsize $\frac{\srm}{\mathrm{kpc}} $& \normalsize	$3.9\pm0.2$ (3.7)	\\ 
\hline
Radius of solar [Fe/H]  & $\frac{\rfeh}{\mathrm{kpc}}$ &	$8.3\pm 0.3 $ (8.9) \\
Enrichment exponent  & $\gfe $ &	$0.19 \pm 0.03$ (0.212)  \\
Metallicity gradient  & $\frac{ \dfe}{\mathrm{dex~kpc^{-1}}}$ & $-0.073\pm 0.002$ (-0.078)  \\
Old disk scale-length& $\frac{\rold}{\mathrm{kpc}}$ & 	$1.3 \pm 0.9$  (2.2) \\
Vertical adaptive scale &  $ a_z $ & $0.85\pm 0.03$ (0.79)
\enddata 
\tablenotetext{a}{Numbers in parentheses are fits using the Bayestar19 extinction map in 240 fields and 7600 stars as a check described in subsection \ref{subsection:test_dustmap}.}
\end{deluxetable}

\vspace{5mm}
\subsection{Inside-out Growth}

We quantify the radial dependence of the star formation history with $x = 0.75$ for the Milky Way's low-$\alpha$ disk. The resulting star formation history is illustrated in Figure \ref{fig:SFH_bestfit}. 
According to this fit, the star formation processes that generated our data set started about $\tau_{m} = 7.5$ Gyr ago. This maximum age differs from the measured age of the oldest star in the data set because (1) a fraction of old stars constrain the old component of the model more than the young one, and (2) this parameter is in units of the dummy variable `true age' of the model, which is convolved over age uncertainties. Therefore, $\tau_m$ is different from a best fit that would not account for age uncertainties, where it would take the value of the oldest star of the `young sample' (where the `young sample' would be defined as the 95\% youngest stars of the APOGEE sample because of our split in young and old components, see Section \ref{subsection:old_component}).

The best fit value of $x$ implies that the star formation rate has been constant at $\sim 10.5$ kpc, a decreasing function of time in the inner $\sim 10.5$ kpc, and an increasing function if time in the outer $\sim 10.5 $ kpc. This means that the Milky Way disk is still forming stars, with a slower decay in the outer disk than in the inner disk. The star formation history decayed slowly in the Solar neighbourhood.

\subsection{Other Parameters}

Even if the present work focuses on the radius-dependent star formation history of the Milky Way disk, the processes determining the evolution of the Milky Way are multiple and complex. We therefore constructed a global model to account for the effect of several of them, in particular \rom~and chemical evolution.

We find a \rom~strength of about $\sigma_\mathrm{RM} = 3.9 \mathrm{~kpc}\sqrt{\tau/7~\mathrm{Gyr}}$. This value is greater than, but consistent with, the amount of radial migration in \cite{frankel_etal_2018} who found $\sigma_\mathrm{RM} \approx 3.4 \mathrm{~kpc}\sqrt{\tau/7~\mathrm{Gyr}}$. The difference results from the different age sets used between that work and the present study. The current age catalog is, on average, younger. So the stars have had less apparent time to migrate the same distance, which pushes the \rom~strength up accordingly. The effects of different age catalogs is discussed more extensively in Sections \ref{subsection:age_sets} and \ref{section:limitations}.

\subsection{Tests}

We test the robustness of the results to various numerical approximations and data uncertainties, and discuss the most important aspects below.

\vspace{6mm}
\subsubsection{Basic Tests} \label{section:tests}
We perform a series of simple tests to ensure that (1) model prediction in data space compare well to our data set, (2) a density model recovers similar work in the literature, (3) integrals are computed with the necessary level of accuracy, (4) the optimization scheme recovers parameters correctly.

\begin{figure}
    \centering
    \includegraphics[width=\columnwidth]{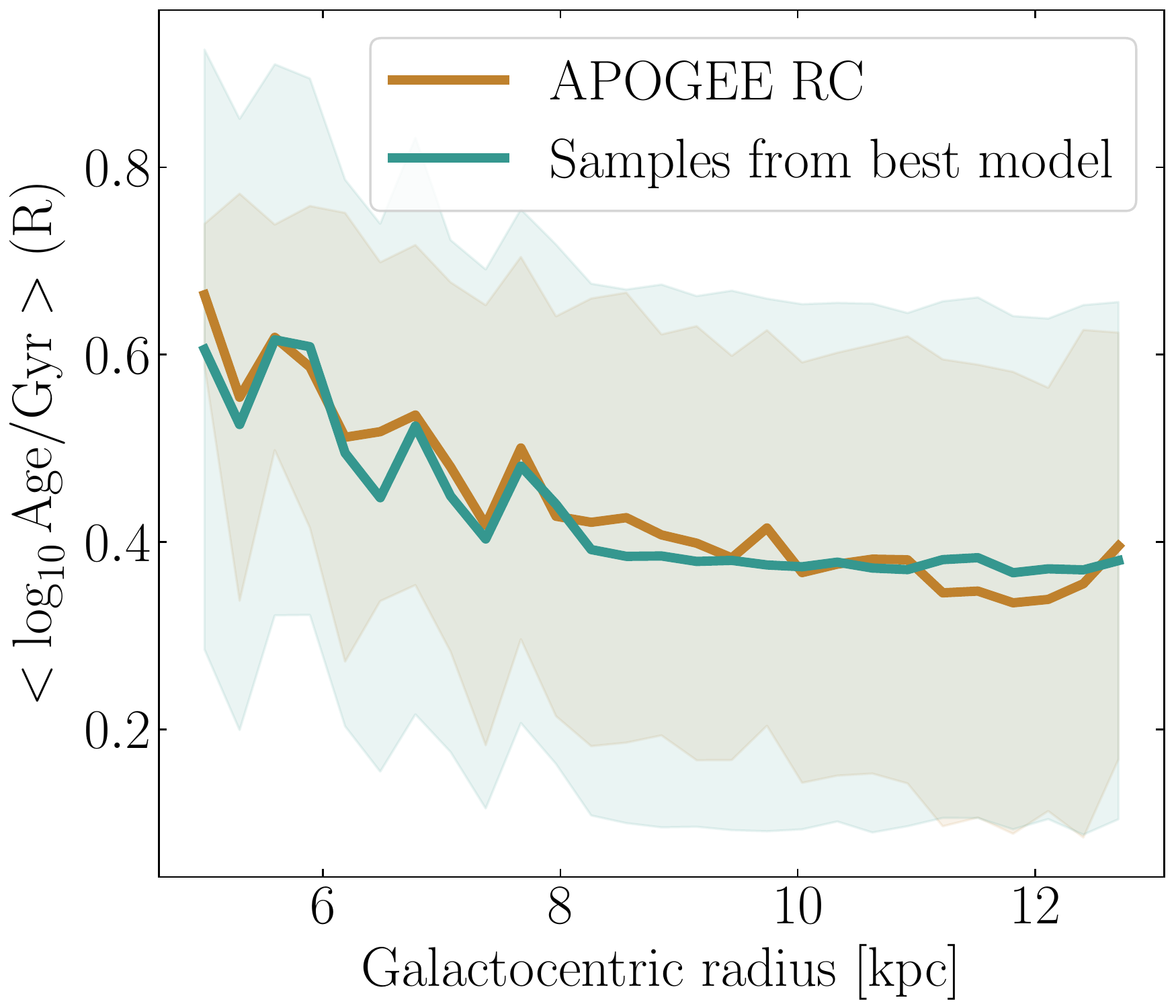}
    \caption{Radial profile of the stellar age distribution in the Galactic disk. The observed data are shown brown, and samples from the best fit model, after applying the APOGEE selection function, are shown in green; the thick lines reflect the mean age and the filled areas are the age dispersion. The jaggedness of the median age is a consequence of the spatially complex selection function, not an indication of spatially discontinuous star formation. This is why the model (green), combined with the APOGEE selection function, is able to reproduce these features even with a smooth star formation history. This Figure illustrates that the observed, present-day age distribution as a function of radius, can be modelled well by our model.
    }
    \label{fig:age_profile}
\end{figure}

\begin{figure}
    \centering
    \includegraphics[width=\columnwidth]{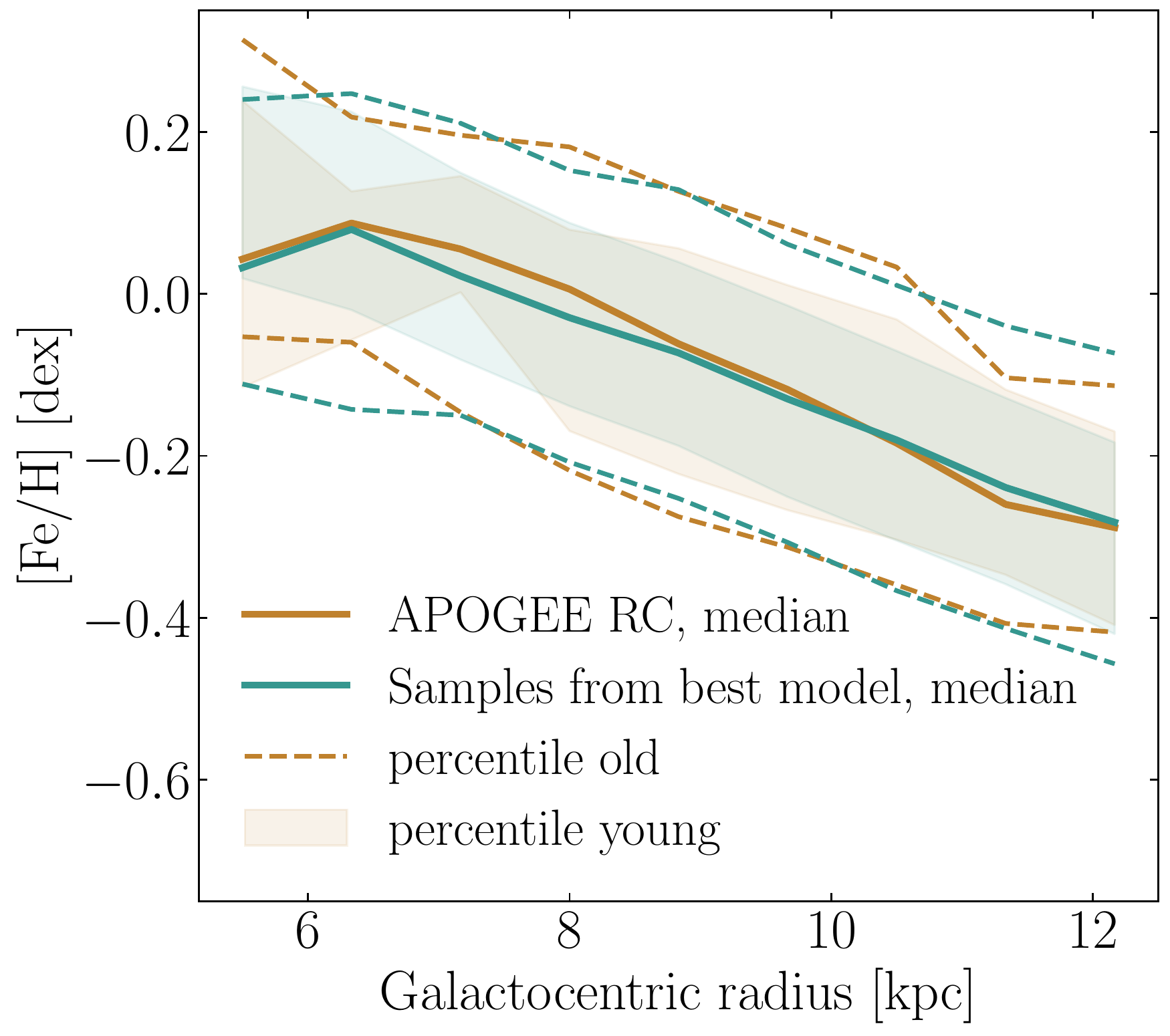}
    \caption{Radial metallicity profile of the observed data (brown) and samples from the best model (green), analogous to Fig.\ref{fig:age_profile}. The solid lines are the median metallicity profiles using all stars, and using a best fit metallicity gradient of $-0.073~\mathrm{dex~kpc^{-1}}$ (see Table \ref{table_best_fit}). The 16\% and 84\% percentiles of two age bins are also plotted: the shaded area for stars younger than 1.5 Gyr, and the dashed lines for stars with ages $2.5 < \tobs < 5 $ Gyr. The metallicity scatter increases with age due to \rom, and seems well reproduced by the model. The slight differences in the inner disk are expected because there are few data points at radii less than 7 kpc. Our model reproduces well both the metallicity gradient and the age-dependent metallicity scatter.}
    \label{fig:metallicity_profile_data}
\end{figure}

\begin{enumerate}
    \item To verify that the fitting procedure has worked, we produce a mock data set from the best fit parameters using the APOGEE selection function and compare it to the real data. The mass-weighted age profile is well recovered, but the predicted age scatter is greater than observed (see Fig.~\ref{fig:age_profile} and Section \ref{subsubsection:model_shortcomings}). 
    Overall, the best fit evolution model reproduces well the observed scatter in age-metallicity at given Galactocentric radius, as shown in Fig.~\ref{fig:metallicity_profile_data}. The mean metallicity profile is well reproduced, indicating that our parametric prescription for the gradual enrichment of the gas was well fitted. In Figure \ref{fig:metallicity_profile_data}, there is an apparent flattening of the metallicity profile at $R < $ 7 kpc, both in the model and in the data. In the present work, these are purely due to selection effects, since our chemical evolution model has a constant metallicity gradient in space and in time. The inner disk APOGEE fields tend to point to high Galactic latitudes away from the mid-plane (Fig.~\ref{fig:selection_function}). Since there are vertical metallicity gradients rising from vertical heating of older, metal-poorer stars, the data set lacks the metal rich stars from the inner mid-plane, underrepresented due to the spatial selection. Of course, this is not an issue in our case because these effects are fully accounted for in our model. But it highlights the importance of accounting carefully for selection effects when using data from surveys to draw conclusions on the evolution of the Galaxy.

    \item We set out a model for the 3D density of the Galactic disk $p(X, Y, Z ~|~ \fe, \tau)$, as in \cite{bovy_rix_etal_2016, mackereth_etal_2017}. Qualitatively, fitting for the scale-lengths and heights in age-metallicity bins gave consistent results with \cite{mackereth_etal_2017}, where at given age, the distribution of metal-rich stars peak in the inner disk and those of metal-poor stars peak in the outer disk. And at given metallicity, the distributions of old stars are in broader `donut-like' structures than those of young stars.

    \item We verify that the normalizing integrals of our global model were evaluated with enough accuracy. Varying the ranges and regularity of the integration grids, we find that increasing the accuracy further does not influence the best estimate results presented here.

    \item To ensure that the optimization scheme (Nelder-Mead algorithm) does not get stuck in local maxima, we optimize the likelihood several times, with different initial parameters.
    Additionally, we check that the optimization scheme was able to recover the true parameters. We sample 5400 mock data points with various parameter values (as well as the best fit) and add noise to their ages from our noise model, Eq.~\ref{eq:noise_model}. We fit these data, and recover the parameters with good precision. The enrichment parameters used to tag birth radii, and the \rom~strength are recovered to 4\%, $\tau_m$ is recovered to 0.5\%, $x$ to 7\% and \tsfr~to 15\% (depending how much noise is added to the ages, here for $\sigma_{\log_{10}\tau} = 0.15$ dex). The scale-length of the disk is recovered to 3\% and $a_z$ to 0.9\%.

\end{enumerate}

\subsubsection{Modelling Extinction \label{subsection:test_dustmap}}

We verify the consistency of the effective selection function of this work (Eq. \ref{eq:selection_function_dust}), using the Bayestar17 extinction map \citep{green_etal_2018}, with the more recent extinction map Bayestar19 \citep{gren_etal_2019}. Since we have excluded all the APOGEE fields where the median extinction of the APOGEE sample is greater than $A_H=0.6$, the Bayestar17 map is largely sufficient for our description of the disk in the 142 remaining disk fields and yields the same effective selection function as the Bayestar19 map. However, the effective selection function differs significantly in the Galactic mid-plane where there are more extinguished fields, which we have excluded. Assuming we can trust the Bayestar19 dust map for such fields, we have fitted the parameters again using the 7600 stars present in the 213 disk fields including those high extinction fields. The best fit parameters are listed in parentheses in Table \ref{table_best_fit}. We find a similar estimate of the disk scale-length \Ro, which is the parameter that would have been most affected by variations in the effective selection function (direct density modelling). The star formation time-scale \tsfr~changes significantly after including more stars from the inner disk (due to co-variances with the scale-length of the old disk component and the fact that the data set has changed but not our outlier fraction $\epsilon$), but the inside-out growth parameter $x$ remains strongly positive.

\subsubsection{Modelling Age Uncertainties}
Stellar ages are challenging to determine. They are modelled quantities rather than direct observables, so using stellar ages relies on assumptions in the underlying stellar evolution models. Ages are heteroscedatic, their uncertainties not well quantified, and their systematic behaviors unknown. In the present work, we assume that asteroseismic ages are the ground truth, and that age errors arise when mapping from stellar spectra to asteroseismic ages through data-driven methods. For the data set we have used, these errors are approximately constant in log age, and the standard deviation of log age on a test set (APOKASC2) is roughly 0.15 dex in decimal logarithm, \citep{ting_rix_2018}. We used a noise model based on this value, where the measured log age is normally distributed around the true value with a 0.15 dex spread (Eq.~\ref{eq:noise_model}). This assumes that the neural network used to map stellar spectra to log age produces Gaussian errors of 0.15 dex. 

To benefit from the direct use of stellar ages, we need to test and understand the impact of the (unknown) uncertainties on our conclusions. We test the robustness of the results to the assumed noise model by varying the value of this scatter between 0.12 and 0.17 dex, and optimizing the likelihood with these new values. We find noticeable changes in the total age of the low-$\alpha$ disk (parameter $\tau_m$) and a (weak) dependency on the inside-out growth parameter $x$ that takes values between 0.6 (underestimate errors) and 0.82 (overestimate errors). 
This leaves our conclusions on inside-out growth unchanged (with a strongly positive $x$), but it affects the comparisons between model predictions and the data set for old stars. So the best fit parameter $\tau_m$ should not be seen as the age of the Milky Way disk, but rather as the maximum age at which we trust our model - based on the age scale and assumed uncertainties of our data.

\subsubsection{Different Age Determination Methods \label{subsection:age_sets}}
In the era of large surveys and data-driven methods, there are at present multiple catalogs of parameters derived for the same stars in large homogeneous data sets. The values inferred from different methods have systematic differences, which are sometimes significant. It is central to realize how these systematics influence our understanding of Galaxy evolution. We test a total of five age catalogs (described in details in Appendix \ref{appendix:age_sets}), on the exact stars that we presented in Section \ref{section_data} (with some minor loss during the cross-match), keeping metallicity and position the same and changing only the age data column in the likelihood optimization. As expected, the best fit values changed between age sets, see Table \ref{tab:several_ages} where the best fit parameter $x$ is summarized, as well as the Galactocentric radius at which the star formation rate is a constant in time $R_\mathrm{const}$. The first data (which we will refer to as `Ting18') set was used for the analysis described above. The ages of the other sets were derived either using \textit{The Cannon} \citep{ness_etal_2015, ness_etal_2016, ness_etal_2019} or a combination of data-driven mapping from abundances to ages and stellar evolution models \citep{das_sanders_2019, sanders_das_2018}. and are named from these references (see Table \ref{tab:several_ages}).

\begin{table}[]
    \centering
    \begin{tabular}{ccc}
         Age set & $x$ & $R_\mathrm{const}$\\
         \hline
         Ting18 &  0.75 & 11 kpc\\
         Ness16 &  0.62 & 12 kpc\\
         Ness18 &  0.50 & 16 kpc\\
         Ness19 &  1.1  & 8 kpc\\
         Sanders18 & -0.3 & --\\
         \hline
    \end{tabular}
    \caption{Inside-out results using different ages}
    \label{tab:several_ages}
\end{table}

These five data sets, composed of the same stars but with different age estimates, lead to five different best-fit values for inside-out growth $x$ due to systematic differences between methods. These differences in inside-out growth best fit arise naturally from the different radial age gradients present in each data set: the data set containing the strongest age radial gradient (Ness19) leads to the strongest inside-out growth best value ($x = 1.1$), and the data that show the weakest age gradient (or the least negative, Sanders18) also has the least positive inside-out growth ($x = -0.3$). Apart from the Sanders18 age set, all data are consistent with an inside-out growth formation scenario for the Milky Way disk,  with some slight difference on the decay of star formation with Galactocentric radius. We found that the Sanders18 age set might over estimate stellar ages at large distance: this is the only method making direct use of Gaia parallaxes, and comparing these ages to the four other spectroscopic ages (with no known bias with distance) shows systematic differences as a function of distance. The systematic zeropoint issues with the Gaia parallaxes are now well studied \citep{leung_bovy_2019} which bias the ages via the the strong mass-luminosity degeneracy for giant stars. Folding this distance bias with the APOGEE spatial selection function can produce an inversion of the overall age gradient in the data set.

\subsubsection{Comparisons with Literature Results}

The test presented above consists in comparing inference results from the same stars, of the same population (red clump), but using ages derived from different methods. In complement, we now use the best fit parameters from these five age sets to make predictions that can be compared with literature. We compare the model predictions to the age distributions from \cite{xiang_2018}, who measured ages from turn off and sub-giant LAMOST stars \citep{cui_LAMOST_2012} with a 0.12 dex precision on log age. They corrected for selection effects, and derived selection-corrected age histograms in Galactocentric radius bins. To compare, we simulate this procedure. We sample stars of all populations (not just red clump) from our best fit evolutionary model and select them in Galactocentric radius bins as in \cite{xiang_2018}. We then add 0.12 dex of noise to the ages, to emulate their age histograms. We find that the best fit parameters obtained from the Ting18 sample (the ages we have focused on during the analysis) are more consistent with the trends in \cite{xiang_2018} than those from the other age sets, at young ages (where it matters and where the data are constraining). The model predictions and those of \cite{xiang_2018} differ significantly at large ages: we under predict the number of old stars. This could come from that our model is mostly constrained by young stars, and that we only model the low-$\alpha$ disk, whereas \cite{xiang_2018} derived these age histograms considering all stars.

\subsubsection{Inside-out Model Variants}

We test several models for the distribution of $\Sigma(\Ro, \tau)$ in order to (1) see whether the general result `the Milky Way disk grew from inside-out' is robust to model variations and to our definitions, and (2) set out to understand what the best description for inside-out growth is. We describe two of them below.

\paragraph{(1) Time-varying Disk Scale-length}
We first build a model of a birth surface density profile with a time varying scale-length, where
\begin{equation}
    \Sigma(\Ro, \tau) = \mathrm{SFH}(\tau) \Sigma(\Ro ~|~ \tau),
\end{equation}
inspired by that in \cite{frankel_etal_2018}.
The star formation history is simply a universally-decreasing exponential, where stars started forming 12 Gyr ago on a time-scale \tsfr, and $\Sigma(\Ro ~|~ \tau)$ is an exponential profile of the disk with a varying scale-length $R_d(\tau) = R_{d,0}(1 - \arexp \frac{\tau}{8 \Gyr})$. The fitted parameter quantifying inside-out growth is \arexp. The fit results in overly strong inside-out growth, \arexp~going to 1 (which confirms and amplifies the inside-out growth results found here). But this model is conceptually deceiving: integrated over time, it predicts an overall disk with a profile that deviates significantly from the expected exponential for disk galaxies. Additionally, this model is restrictive: by construction, it is (1) unable to describe outside-in growth in a physically plausible way, (2) unable to describe very strong inside-out growth, which would result in stars born with a negative scale-length.

\paragraph{(2) Radius-varying Star Formation Peak}
We test another model, where inside-out growth is not described by a radially-varying star formation time-scale as in equation \ref{eq:SFH_def}, but rather with a radially-dependent star formation peak. This raises the question of what inside-out growth is: do stars form on a longer time-scale at large radii, or does star formation begin later at large radii? Unfortunately, the second question, `when do stars begin to form?' requires data that are constraining at large ages, which are currently not available due to the red clump selection. The best fit parameters are not in contradiction with inside-out growth and samples from that model reproduce the observed data better than the model we chose to show in this analysis. But the fitting procedure is unfortunately not more conclusive in a quantifiable way. This is because age errors at large ages are so large, and old data are so sparse, that our likelihood function is almost flat in the parameter quantifying the radial dependency in the star formation peak.

%
%
%
%
%
%


\section{Astrophysical Implications}\label{section:astrophysical_implications}

\begin{figure}
    \centering
    \includegraphics[width=\columnwidth]{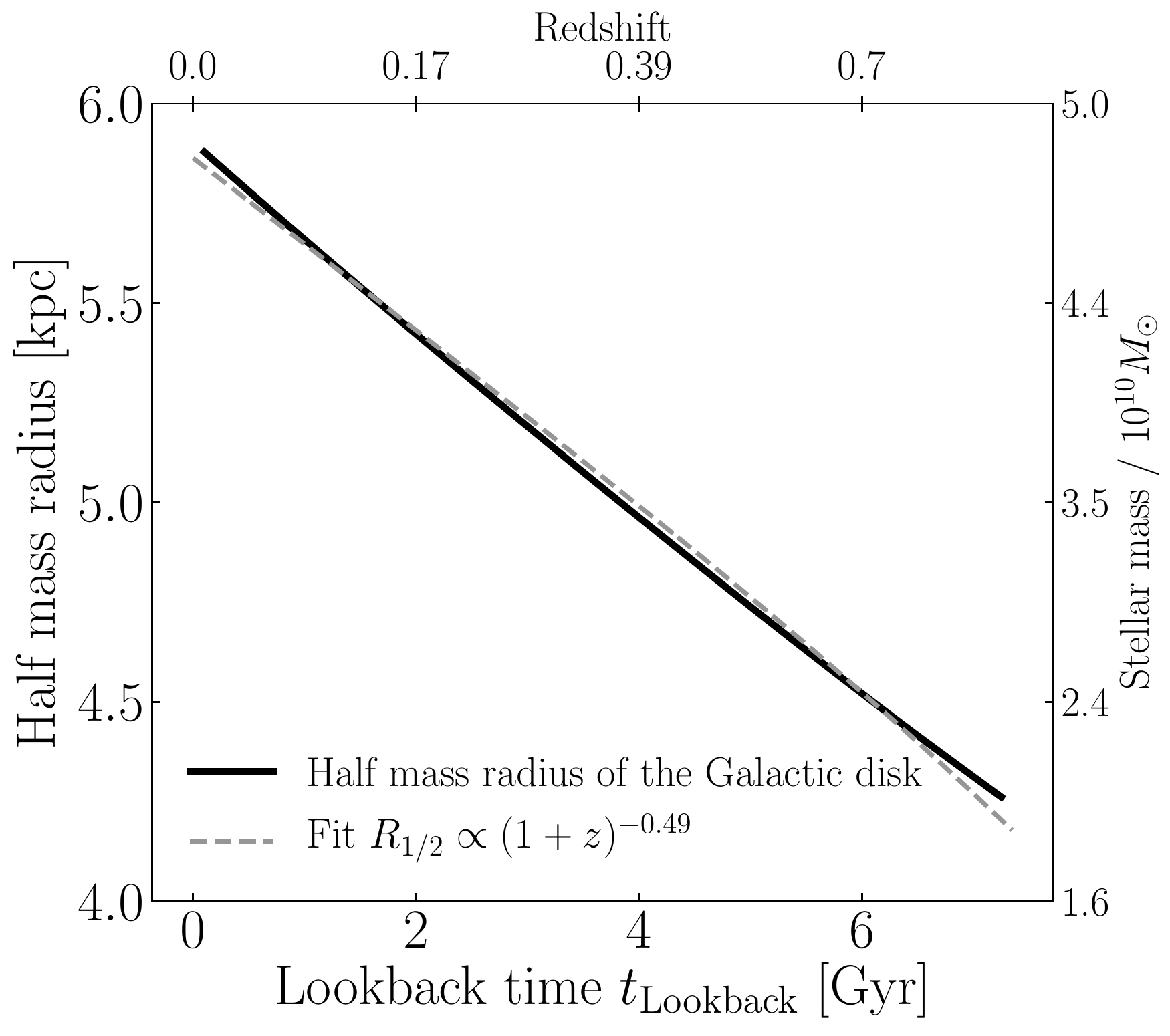}
    \caption{Size evolution of the Milky Way's low-$\alpha$ stellar disk over the past 7 Gyr, according to the best fit model (as computed in Eq.~\ref{eq:half_mass_radius}). The disk has grown by 43\% its size 7.5 Gyr ago.
    The dashed grey line is an approximation of this evolution by $R_{1/2} \propto (1+z)^\alpha$ with $\alpha = -0.49$. The stellar mass as a function of lookback time (as computed in Eq. \ref{eq:M_star_disk_evol}) is illustrated on the right hand side y axis.}
    \label{fig:half_mass_radius}
\end{figure}

We have constructed and constrained a model that describes how the Milky Way's low-$\alpha$ disk may have built up and grown over time. We discuss below the implications in the more general context of the formation and evolution of galactic disks, and show that under the assumptions of this work, the Milky Way is typical in this aspect.

For external galaxies, there exist two families of observational constraints for inside-out growth. Redshift-size relations are obtained by measuring the size of the visible disks (half-light radius, or effective radius), assuming exponential or de Vaucouleur \citep{deVaucouleur_1948} profiles. These observations reflect the overall size of galaxies of fixed stellar mass (derived from their observed light) at given lookback times, and show that on average, disks were smaller in the past. Since these observations are made on populations of galaxies of a given stellar mass at all redshifts, they do not reflect the time evolution of individual galaxies (especially as, due to star formation, the stellar mass of a galaxy should in principle increase with time). In this picture, a star-forming disk galaxy's stellar mass and size should grow with time. Interpreting the evolution of scaling relations with time requires modelling \citep[e.g.,][]{mo_mao_white_1998}.

On the other hand, galactic archaeology approaches have studied the present-day positions of different populations of stars in individual galaxies, including the Milky Way. But stellar ages have large uncertainties, and present-day stars Galactocentric radii are not their birth radii. If stellar ages indicated the lookback time at which stars formed, and if present-day stellar positions reflected well their birth positions, this approach would provide a direct knowledge of the formation of galactic disks. But under extensive \rom~and diffusion, information on stellar birth sites, and therefore local star formation histories, is lost. 

Measured mass weighted age gradients are small. For instance, \cite{goddard_2017} found mass weighted age gradient on average consistent with zero. This is in apparent contradiction with the picture provided by redshift-size relations, implying that old stars should be more centrally concentrated if stars were not radially redistributed in some way. If these differences are physical (and not due to measurement systematics), two scenarios can reconcile these observations: (a) either star formation is constant across disks, and disks grow in size due to redistribution of stars or mergers for non isolated galaxies, or (b) star formation moves outwards with time but \rom~erases the traces of it, such that present-day studies only see the weak remnants of formation gradients. Our present model for the Milky Way disk allowed both pictures to exist through the parameter $x$, that would be zero in (a) and positive in (b). The data preferentially constrained it towards (b) since the best fit $x = 0.75$.

Below, we use the best fit model to predict (1) the evolution of the half-mass radius of the Milky Way disk with time, (2) the evolution of its surface brightness profile and its half-light radius, to compare to redshift-size relations, and (3) the present-day positions of stars of different ages.

\subsection{Evolution of the Disk Half-mass Radius}
We use the best fit model to predict the size of the Milky Way disk at different times in the past. Accounting for inside-out growth and radial redistribution of stars, we compute the radius containing half of the total disk mass,
\begin{equation}
M(< R_{1/2}) = M_{tot}/ 2,
\end{equation}
as a function of lookback time $R_{1/2}(\tlookback)$ by solving numerically for the radius  $R_{1/2}(\tlookback)$ enclosing half of the total disk stellar mass $M_{tot}$.

The stellar mass $M(< R_{1/2})$ contained within $R_{1/2}$ is defined as
\begin{equation}\label{eq:half_mass_radius}
\begin{split}
M(<& R_{1/2}, \tlookback) =  M_{tot}\int_0^{R_{1/2}} \int_0^\infty \int_{\tlookback}  ^{\tau_m}p(\Ro)\\
&\times p(\tau~|~\Ro)p(R ~|~ \Ro, \tau - \tlookback) \mathrm{d}\tau \mathrm{d}\Ro \mathrm{d}R,
\end{split}
\end{equation}
where \tlookback~is lookback time, $\tau$ is time from now (dummy lookback time in the integral, marking the birth of stars) and the quantity $\tau - \tlookback$ is the age of stars at lookback time $\tlookback$ and is used to compute the radial migration term $p(R~|~\Ro, \tau - \tlookback)$. The evolution of the half-mass radius for the best fit model is illustrated by the black solid  line in Figure \ref{fig:half_mass_radius}.
The size of the disk has evolved almost linearly in time from about 4.2 kpc 7.5 Gyr ago to about 5.9 kpc today. This implies an almost constant growth rate of $\dot{R}_{1/2} = 0.2~\mathrm{kpc~Gyr^{-1}}$. The associated time-scale of radial disk growth is $\tau_{R} = {R}_{1/2}/\dot{R}_{1/2} = 30 $ Gyr. This compares well with the findings of \cite{pezzulli_2015}, who measured $\dot{R}_{1/2}$ using star formation and $R_{1/2}$ assuming an exponential surface density profile of about $30$ disk galaxies and quote the same growth time-scale. There may be differences by factors of a few due to systematic uncertainties or the definition of a disk size used (half-mass radius, half-light radius, or exponential scale-length).

Using $\Omega_\Lambda = 0.7$, $\Omega_m = 0.3$, $H_0 = 70~\mathrm{km~s^{-1}~Mpc^{-1}}$, the half-mass radius obtained from Eq.~\ref{eq:half_mass_radius} is well approximated by a function of redshift $z$ as $R_{1/2} \propto (1+z)^\alpha$, with $\alpha= - 0.49$. This seems globally consistent with redshift-size measurements of external disk galaxies. \cite{franx_etal_2008} find an average relation for galaxies of stellar mass $M_\star > 2.5\times 10^{10} M_\odot$ with $\alpha = -0.6 \pm 0.1$ and \cite{trujillo_2006} find $\alpha = -0.40 \pm 0.06$ for late type galaxies of stellar mass $M_\star > 3\times 10^{10} M_\odot$. This approximation of the half-mass radius is illustrated by the grey dashed line in Figure \ref{fig:half_mass_radius}. However, this direct comparison is only approximate: the literature relations were fitted for galaxy populations of given stellar mass, whereas this work predicts the evolution of the half-mass radius of the Milky Way at different times of its evolution, including its growth in stellar mass. We propose to compensate for these differences in the next subsection.

\subsection{Evolution of the Half-light Radius: Comparing to Redshift-Size Relations}

\begin{figure}
    \centering
    \includegraphics[width=\columnwidth]{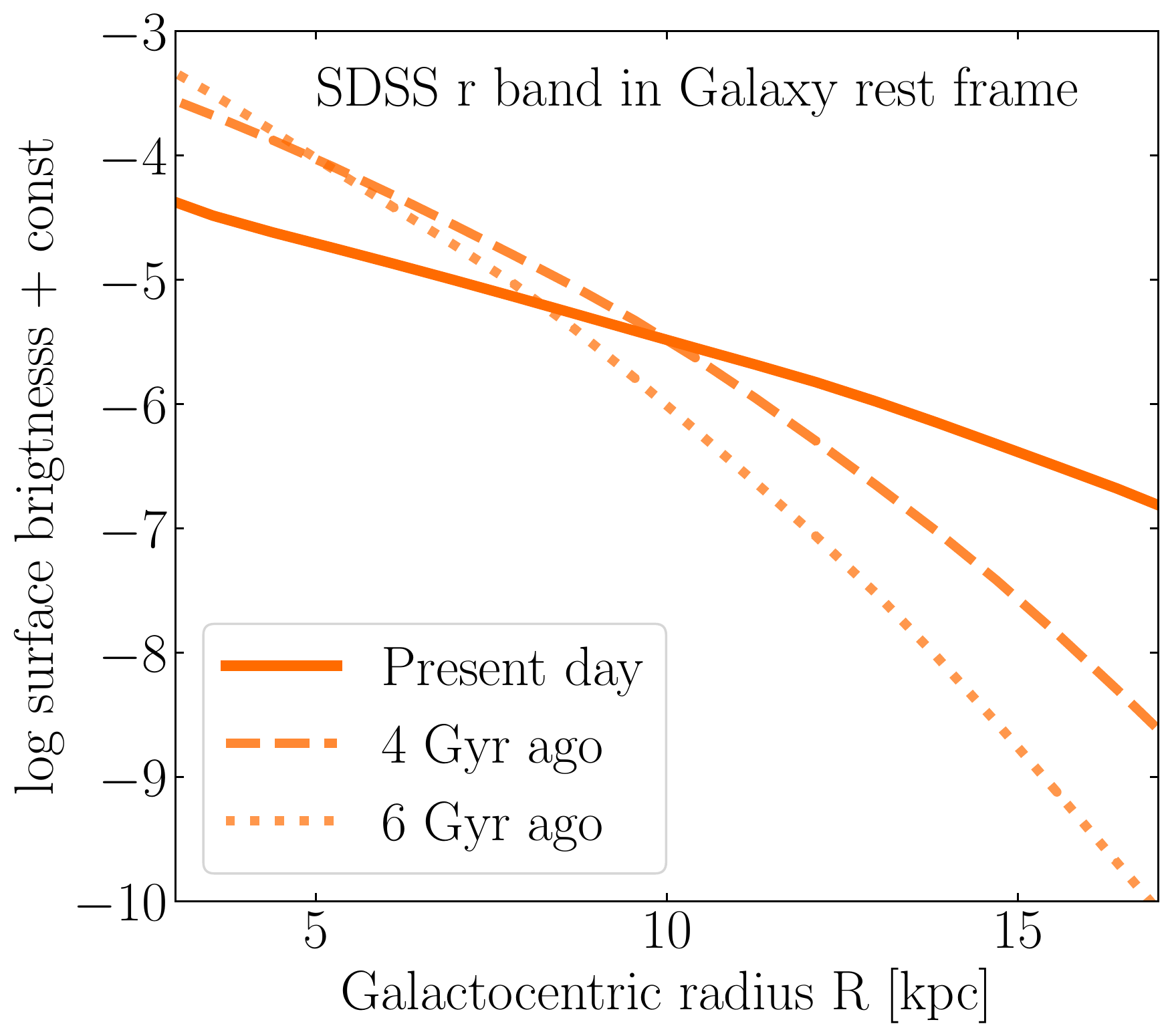}
    \caption{Predicted SDSS r band surface brightness profile for the Galactic disk at three different lookback times. This illustrates the inside-out growth of the stellar disk implied by our model fit.}
    \label{fig:surface_brightness}
\end{figure}

\begin{figure}
    \centering
    \includegraphics[width=\columnwidth]{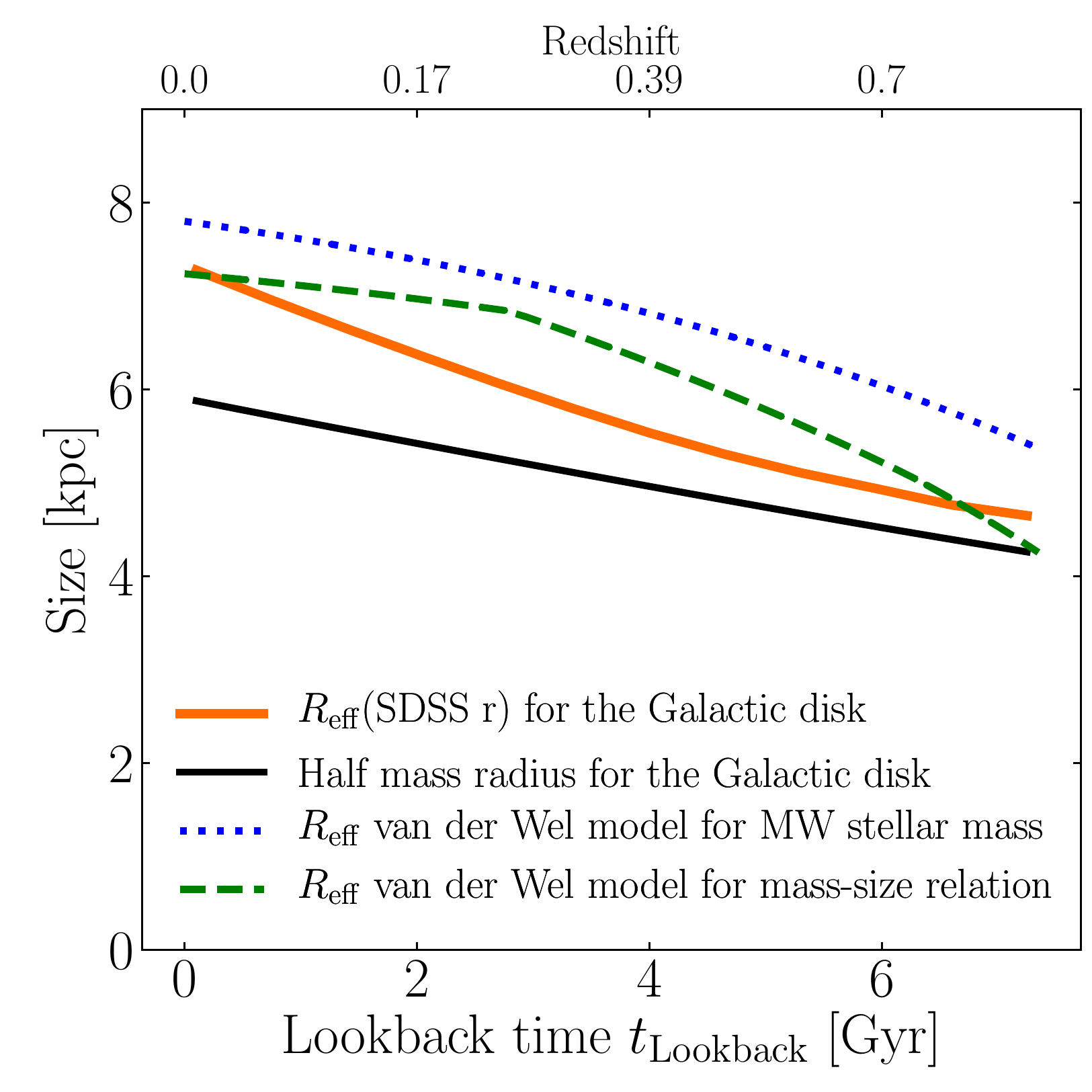}
    \caption{Predicted half-light radius as a function of lookback time for the Milky Way disk (thick orange), compared with that of other galaxies (green and blue). The blue dotted line is a prediction from the \cite{van_der_wel_etal_2014} model for Milky Way mass redshift-size relations, as an element of comparison. It is different from the evolutionary track of a single galaxy, because galaxies grow in stellar mass and size (whereas the \cite{van_der_wel_etal_2014} blue line is at fixed stellar mass). The green dashed line shows the the same prediction of the \cite{van_der_wel_etal_2014}, but including the evolution of the stellar mass of the Galaxy as in Eq.~\ref{eq:M_star_disk_evol}. It leads to a redshift-size evolution close to our inference for the Milky Way.} We include the half-mass radius (from Fig.~\ref{fig:half_mass_radius}) in black as a reference. 
    \label{fig:half_light_radius_evol}
\end{figure}

In external galaxies, the physical half-mass radius is not a direct observable. It is often assumed that stellar light at different wavelengths traces stellar mass of different populations, and the half-light radius is used as a proxy for the disk size. We predict the surface brightness profile and the half-light radius that would be measured at different lookback times and in different photometric bands, roughly emulating observations of the Milky Way at different redshifts. For this purpose, we use grids based on the single stellar population synthesis code E-MILES SEDs \citep{vasdekis_2010, vazdekis_2012, ricciardelli_2012} to predict the mass-to-light ratio of a single stellar population of given age and metallicity. Assuming a Kroupa initial mass function \citep{kroupa_2001}, we use the best fit model to predict the distributions of $p(\tau, \fe, R ~|~\ppm, \tlookback)$ at given lookback time to estimate the stellar mass density at radius $R$ (up to a normalization constant). The resulting surface brightness profiles in the SDSS-r band are plotted as an example in Figure \ref{fig:surface_brightness}. Qualitatively, the trends are similar in other bands. Generally, blue bands are spatially more extended and red more centrally concentrated. This is expected because shorter wavelengths are better tracers of young populations.

From these surface brightness profiles, we solve a similar equation to Eq.~\ref{eq:half_mass_radius} for the half-light radius, illustrated by the thick orange line in Fig.~\ref{fig:half_light_radius_evol}. The half-light radius is greater than the half-mass radius, in particular at late times, when the young stars dominating the light are spatially more extended than the overall stellar population. This compares well with the model of \cite{van_der_wel_etal_2014}, who fitted a model to galaxies at different redshifts (illustrated in dotted blue in Fig. \ref{fig:half_light_radius_evol} for galaxies of stellar mass $M_\star \approx 5\times 10^{10} M_\odot$). However, this only serves as an element of qualitative comparison. This model fitted the size of galaxies of $\sim$ Milky Way stellar mass at different redshifts. But the stellar mass of the Milky Way has grown with time. Therefore, the comparison holds best at $z=0$. At higher redshifts, one should compare to the sizes of galaxies of smaller stellar mass, which should also be smaller than the blue dotted line that Fig.~\ref{fig:half_light_radius_evol} predicts.

\cite{van_der_wel_etal_2014} have also fitted the evolution of the total stellar mass $M_\star$ - size $R_\mathrm{eff}$ relation as a function of redshift $z$. With our best fit model of the evolution of the stellar mass of the Milky Way, we can use the \cite{van_der_wel_etal_2014} fits to predict the corresponding size. 
Assuming that the total stellar mass of the Milky Way is the sum of the bulge mass $M_b$ and that of the disk, and assuming that the bulge mass has been constant over the past 8 Gyr (because the bulge is relatively old compared to the disk \citep{bland-hawthorn_gerhard_2016}), we consider the total Milky Way stellar mass as 
\begin{equation} \label{eq:M_star_disk_evol}
    M_\star(t)=M_b + M_d(t).
\end{equation}
We assume a bulge stellar mass of about $1.5\times 10^{10} M_\odot$ \citep{bland-hawthorn_gerhard_2016} and the fraction of mass in the bulge about 30\% at present. The evolution of the stellar mass is illustrated in Fig. \ref{fig:half_mass_radius}.

Following \cite{van_der_wel_etal_2014}, we assume a stellar mass--size relation of the form
\begin{equation}
    R_\mathrm{eff}(z) = A(z) \left( \frac{M_\star(z)}{5\times 10^{10} M_\odot}, \right)^{\alpha(z)},
\end{equation}
where $A(z)$ and $\alpha (z)$ were measured by \cite{van_der_wel_etal_2014} at different redshifts. We interpolate these values linearly as a function of redshift and compute the corresponding value of $R_\mathrm{eff}(z)$ and our predicted Milky Way's stellar mass as a function of time. The resulting size as a function of lookback time (or redshift) is illustrated with the dashed green line in Figure \ref{fig:half_light_radius_evol}, and compares well with our predicted evolution of the half-light radius of the Milky Way.

\subsection{Present-day Scale-lengths of Stellar Populations}

\begin{figure}
    \centering
    \includegraphics[width=\columnwidth]{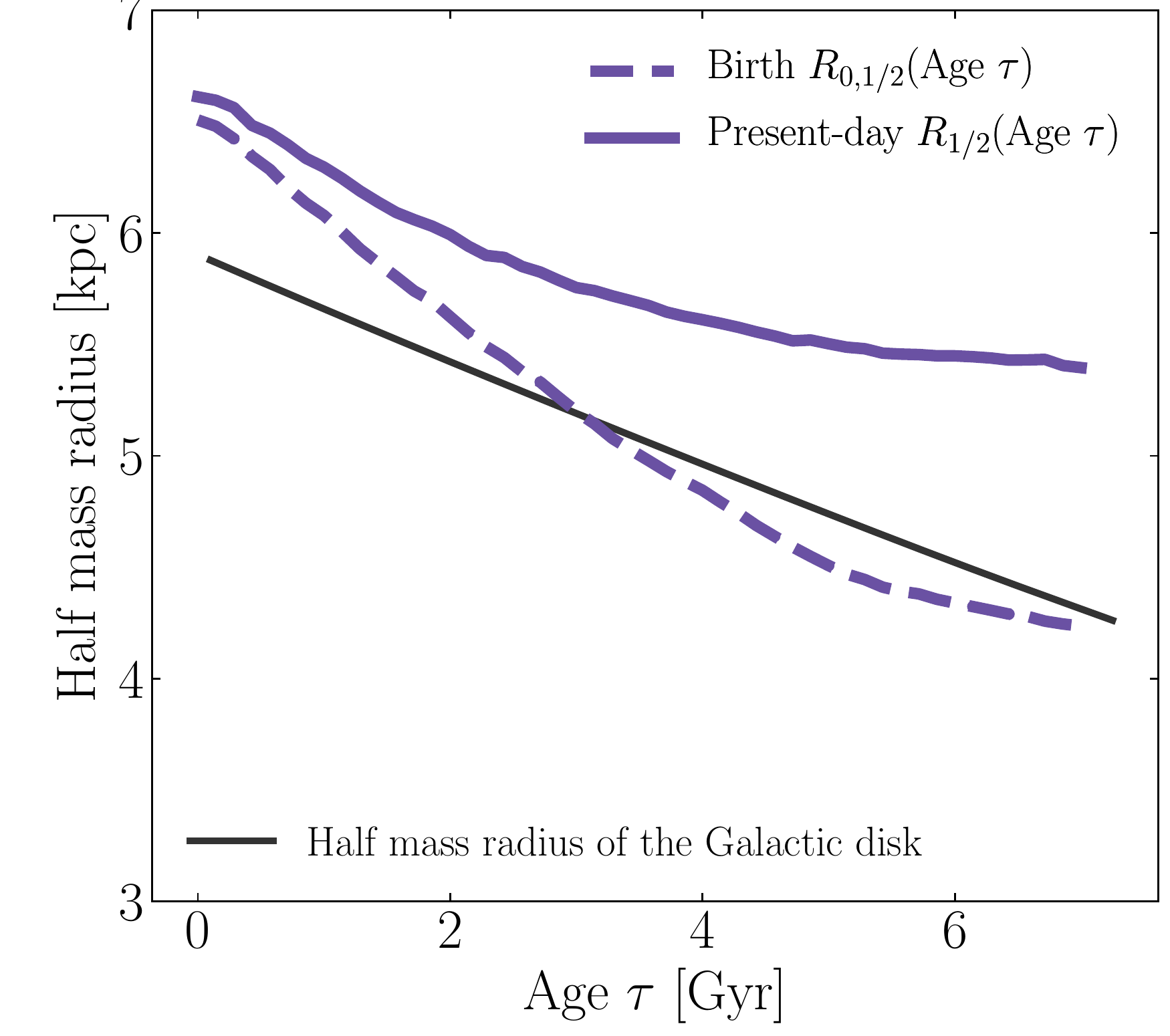}
    \caption{The present-day half-mass radii of stars with ages $\tau$, $R_{1/2}(\tau)$, compared to the star formation scale-length a time $\tau$ ago. We show $R_{1/2}(\tau)$ from the best fit model with solid purple and compare it to the analogous, hypothetical radii expected in the absence of \rom~(dashed purple). As a reference, we overplot the overall size evolution of the disk at different lookback times in solid black. Radial mixing leads to a moderate present-day dependence of $R_{1/2}$ on age, even in the presence of distinct inside-out growth in the star formation history.}
    \label{fig:size_populations}
\end{figure}

In nearby galaxies, it is common to determine the spatial distribution of stars of different populations \citep[e.g.,][]{gogarten_etal_2010}. When old stellar populations are found more centrally concentrated than the young ones, this is interpreted as the result of inside-out growth.
    
We compute these quantities in two cases: (a) for our Milky Way best fit, and (b) for our best fit excluding radial migration (i.e. finding the half-mass birth radii). (b) does not lead to a scale-length that is possibly observable today if stars migrate, but decouples the contributions of the star formation history and \rom~to the growth of the disk. 
To compute these, we perform a Monte Carlo simulation sampling ages, present-day Galactocentric radii and Galactocentric radii at birth for stars from the best fit model. For each sample, we compute the present-day radius $R_{1/2}(\tau \approx \tlookback)$ containing half of the mass of stars at given age, and similarly for the birth radius $R_{0,1/2}(\tau \approx \tlookback)$. These two quantities are plotted in purple in Figure \ref{fig:size_populations}.

We find a significant difference in the half-mass radius of different stellar populations that have migrated (solid purple in Fig. \ref{fig:size_populations}) and those that have not (dashed purple). In the limits of our modelling, we can argue that if we were to observe the Milky Way present-day stellar populations in order to constrain inside-out growth, and assume its stars do not redistribute radially, we would overestimate the scale-length at birth of old stars by about 1 kpc, and therefore underestimate the amount of inside-out growth. This is a caveat for galactic archaeology approaches that cannot be neglected for disk galaxies with significant \rom. For the Milky Way, this can be overcome by detailed modelling, because one can measure a physical scatter in the age-metallicity distributions, and interpret it as an effect of \rom. However, for external galaxies, the question is more challenging as only integrated properties are derived (the mean, but not the scatter). Therefore, the present-day distributions of stellar populations can be a weak diagnostic for inside-out growth, giving only lower limits.

\subsection{Weakening of Age Radial Gradients}

\begin{figure}
    \centering
    \includegraphics[width=\columnwidth]{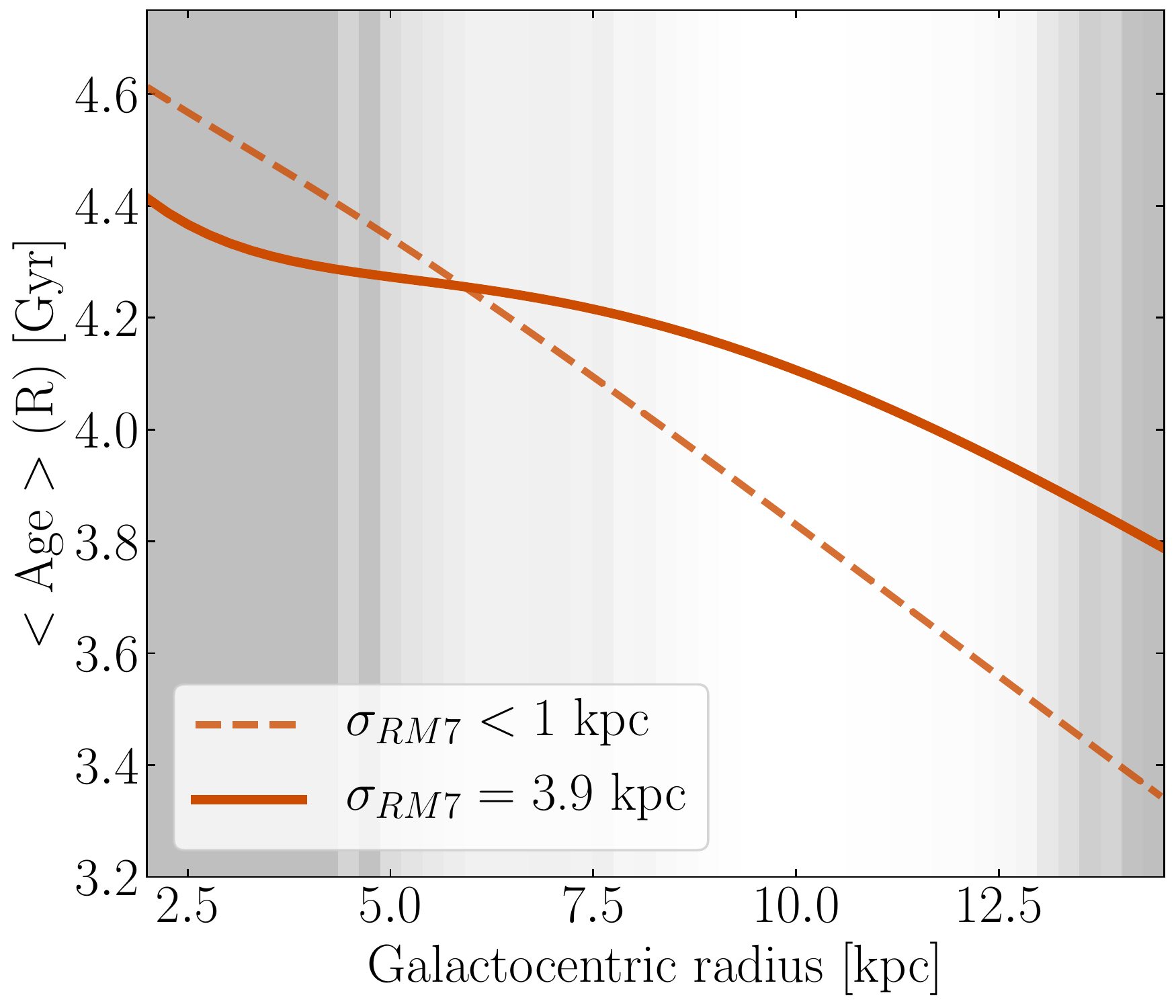}
    \caption{Radial profile of the mean stellar age in the Galactic disk, determined through our model fit (solid line, for $\sigma_{RM7}=3.9$~kpc computed using Equation \ref{eq:age_profile}). This is compared to the hypothetical case of `no \rom' (dashed line), where stars stay at their birth radii (also determined from our model fit). Age gradients resulting from the distinct inside-out growth that the best model implies are severely weakened by \rom. Areas containing few data to constrain our model at their present-day Galactocentric radii are shaded in proportions of the Galactocentric radius histogram in Fig.~\ref{fig:R_histogram} on a logarithmic scale.}
    \label{fig:age_profile_pred}
\end{figure}

Age gradients are expected to result from inside-out growth: if stars formed on shorter time-scales in the inner disk than in the outer disk, then the proportions of old stars should be larger in the inner disk than in the outer disk. But in external galaxies, the mass weighted age gradients seen in MaNGA seem weak for a large number of disk galaxies \citep{goddard_2017, bundy_2015_manga}.

We compute the expected mass weighted age profile from our model and the best fit parameters with
\begin{equation}\label{eq:age_profile}
    \langle \mathrm{Age} \rangle (R) = \iint \tau \cdot p(\Ro, \tau ~|~ R) \mathrm{d}\Ro \mathrm{d}\tau,
\end{equation}
and illustrate it in Figure \ref{fig:age_profile_pred} for two cases: no radial migration (where we take $\srm < 1$ kpc), and our best fit radial migration. We find that, in the `no migration' case, the best fit star formation history produces an age gradient that is already weak, with only about $-0.1 \mathrm{~Gyr~kpc^{-1}}$. But in addition, \rom~weakens severely this gradient. Therefore, it is possible for a galaxy to have grown from inside-out without exhibiting strong age radial gradients, making present-day age gradients a poor diagnostic of inside-out growth.

%
%
\section{Limitations of our Methodology}\label{section:limitations}

Even though the present method is promising for later studies of Galaxy evolution, we state below important shortcomings that should be addressed in the future.

\subsection{Inevitable Limitations: Direct Use of Ages}

The direct use of ages in global models for the Galactic archaeology field is new and can only be as powerful as the ages are good.
Since our method relies \textit{directly} on stellar age measurements, the results are directly affected by any systematics in age determination methods (see Section \ref{section_results}). These systematics are numerous and unknown; and age uncertainties are large and heteroscedatic. Age determination methods differ by their complexity. Data-driven methods depend strongly on the reliability of their training sets, those using luminosity and distances may induce large scale spatial biases, affecting our understanding of Galaxy evolution. To estimate the extent of these limitations, we have considered several data-driven age catalogs, and presented the results for the fit on the Ting18 sample \citep{ting_rix_2018}. Even though this data set has its short comings, such as an unexpectedly small fraction of old stars, it showed more reliability than the three others in the several aspects:
\begin{itemize}
    \item the precision of the determined ages on a test set on APOKASC 2 red clump stars was the highest with 0.15 dex scatter in log age;

    \item model predictions using the best fit on these data compared better than those the three other data sets to independent work on selection-corrected age histograms using LAMOST turn off stars;

    \item the age determination method (spectroscopic ages) has no known dependencies on distance measurement, so there should not be age biases with distance, which is of fundamental importance to study inside-out growth through the spatial structure of ages;

    \item the training of the neural networks that determined these ages was done only on red clump stars, so the neural networks were not learning possibly undesired information from other stellar evolutionary stages.
\end{itemize}
Inevitably, the present results are bound to all assumptions that were made by using this data set and by modelling its age uncertainties.

\subsection{Limitations from the Data Selection}
This work made exclusive use of red clump stars, because they are excellent standard candles, they are bright and they are abundant in the Galactic disk. However, this stellar population is younger than the underlying mean population, with an age distribution that peaks at 2 Gyr and that has few stars older than 5 Gyr (see Fig. \ref{fig:age_distribution}). This introduces two shortcomings. First, most of the information constraining our model lies between 1 - 4 Gyr. Therefore, our model describes best the most recent evolution of the Galactic disk, but not earlier times when the physical processes of disk formation were supposedly faster. Secondly, we corrected for population selection biases by modelling their age distribution (or, more accurately, the relative mass of stars on the red clump evolutionary stage, at given age and assuming an initial mass function). This was based on theoretical expectations of a star's lifetime on the core helium burning stage, which is only constrained by theoretical models of stellar evolution.

In order to constrain the evolution of the disk at earlier times, and reduce stellar population effects on ages, we should turn to a different stellar population with more old stars, for example RGB stars for which distances are known precisely \citep{hogg_eilers_rix_2018, leung_bovy_2019}. But this would introduce more technical difficulties that are not straightforward to overcome: using standard candles limits the number of integrals to evaluate in magnitude-limited selection functions, bringing computational expenses down. However, the lack of old stars in our red clump data set, and the large age uncertainties at large ages, made us unable to build and constrain a more adequate model for the early behavior of star formation. This question is left to be addressed in the future.

Additionally, due to the spatial selection of stars in our survey, most of the stars we have used lie in the outer disk (5-14 kpc, but with few stars between 5 and 7 kpc, see Fig. \ref{fig:R_histogram}). Since these are the data constraining our model, our results describe best the outer disk of the Milky Way. But the inner low-$\alpha$ disk should contain additional information on inside-out growth in its spatial-age-metallicity distributions, unless it had a different formation scenario \citep[as argued in e.g.][]{haywood_2019}. As shown in the test of subsection \ref{subsection:test_dustmap}, including more APOGEE fields with data from the inner disk mid-plane affects some parameters, for example the star formation time-scale \tsfr and the inferred scale-length of our outlier model (old component). We have kept consistency with the data set used in \cite{frankel_etal_2018}, but we plan to turn to APOGEE DR16 and the inner Galaxy in future work.

\subsection{Model Shortcomings}\label{subsubsection:model_shortcomings}

To construct a parametric model for the Galactic disk, we have made several assumptions. The core of this model, a radially-dependent star formation history, was built to satisfy the following properties: (1) be qualitatively plausible; (2) be sufficiently flexible to allow the `no inside-out growth', `inside-out growth' and `outside-in growth' scenarios to exist, and to be fit; (3) be a global decreasing exponential in the case of `no inside-out growth', and (4) be such that our data can constrain it.

The functional form presented in Eq.~\ref{eq:SFH_def} satisfies the properties (2)-(4), and is good with (1) on average. However, it proposes a rather simplistic description of the star formation history: star formation begins at the same time over the entire disk and the only parameter quantifying inside-out growth is the time-scale regulating the star formation. By definition, such a functional form produces an age distribution that has a large variance, and our model does not reproduce exactly the age distribution of the data at given Galactocentric radius. This can be seen in Fig.~\ref{fig:age_profile}, where our model predicts the mean age profile correctly, but overestimates the standard deviation, and is skewed at large ages.
We have experimented with different functional forms for the star formation history, where star formation rises at different times at different Galactocentric radii (see Section \ref{section_results}). Samples from such a model compared much better to the observed data than those from the model we have chosen to present. But due to the lack of old stars in our data set, and large age uncertainties at large ages, we were not able to fit this model reliably (and condition (4) was violated).

In order to improve the description of the data in a physically sensible way, it will be important in the future to introduce realistic parametric models inspired by simulations.

%
%
\section{Summary and Future Prospects}



We have built a model to constrain the degree of inside-out growth of the Milky Way disk. Applied to APOGEE red clump stars, the model fit leads to significant inside-out growth and has several implications. In this modelling context, we find that
\begin{itemize}
    \item the model fit implies a growth rate of the stellar disk of $0.2~\mathrm{kpc~Gyr^{-1}}$ and a present-day half-mass radius of 5.9 kpc, resulting from a 43\% growth over the past 7.5 Gyr;
    \item comparing the half-mass and half-light radii with redshift-size relations and with the evolution of the mass-size relations of other galaxies shows that the Milky Way is a rather typical disk galaxy in terms of present-day mass and size, but also in terms of global mass-size evolution. We fitted the evolution of the Milky Way's half-mass radius  $R_{1/2} \propto (1 + z)^{-0.49}$);
    \item important radial redistribution of stars erases present-day evidence for the past history of the disk, including the spatial variations of the star formation history and therefore inside-out growth. This implies that a local age histogram differs from a local star formation history as local stars may have been born at different Galactocentric radii and stars formed locally may have migrated to different radii;
    \item present-day age gradients can be a poor diagnostic for inside-out growth if \rom~is strong and makes population studies in external galaxies challenging as assessing the strength of radial orbit migration is more difficult (the physical scatter in age-metallicity resulting from radial orbit migration is not accessible through integrated light). Indirect possibilities to assess radial mixing strength in external galaxies and their impact on the stellar age structure are, for example, dynamical studies as in \cite{gogarten_etal_2010}.
\end{itemize}

This methodology builds on a large data set, with extended spatial coverage and for which stellar ages are known, combined with a forward model with parameters physically understandable. This allows to quantify the amount of inside-out growth in the Milky Way disk. However, there are clear limitations to outcome: the data set we have used is overall young (whereas the evolution of the disk was possibly faster at early times), which allows us to fit the recent evolution of the disk well but not earlier times. In addition, the results are bound to all assumptions made while constructing this model.

To confirm and improve these results, reliable age estimates of stellar populations that cover wide age and Galactocentric radius ranges are needed. Red clump stars are excellent distance indicators, but have poor age estimates and are younger than the underlying population. We showed that five different age sets for our red clump stars lead to five different estimates of the degree of inside-out growth, which is a general problem for Galactic archaeology.

With a data set with precise distances \citep[e.g.,][]{hogg_eilers_rix_2018, leung_bovy_2019} and less prone to population selection effects, it should be feasible to improve the inside-out growth model for a better description of the build up of the Milky Way disk and inspire the model construction from numerical simulations, to benefit from direct physical insights. But these requires more computational issues to overcome.

\section*{Acknowledgements}
It is a pleasure to thank Maosheng Xiang and Ted Mackereth for providing us insights on their results for our safety checks, and Gregory Green for practical help on using the Bayestar17 extinction map. We thank Diane Feuillet and Andy Gould for useful discussions on APOGEE selection function, and Alina Boecker and Ryan Leaman for advice on stellar populations. We are grateful to Sofia Feltzing and Paul McMillan for helping to improve the clarity of the writing.
This project was developed in part at the 2018 NYC Gaia Sprint, hosted by the Center for Computational Astrophysics of the Flatiron Institute in New York City and at the 2018 Gaia-LAMOST hack-a-thon, held in Shanghai.
N.F. acknowledges support from the International Max Planck Research School for Astronomy and Cosmic Physics at the University of Heidelberg (IMPRS-HD). H.-W.R. received support from the European Research
Council under the European Union's Seventh Framework
Programme (FP 7) ERC Grant Agreement n. [321035]. YST is grateful to be supported by the NASA Hubble Fellowship grant HST-HF2-51425 awarded by the Space Telescope Science Institute.

The following softwares were used during this research: Astropy \citep{astropy}, Matplotlib \citep{matplotlib}.

\end{CJK*}

\appendix

\section{Inside-out growth inference using different age sets \label{appendix:age_sets}}

As part of the analysis, we explore the effect of age determination methods on the results. We repeat the analysis keeping the same stars, the same [Fe/H] and position columns, but changing the age column using five different age sets described below. We find that the overall inside-out growth scenario is robust to age systematics, but the details can change quantitatively. We describe the detailed procedure and results for each age set below.

\paragraph{Ness16} We use ages derived with \textit{The Cannon} \citep{ness_etal_2015, ness_etal_2016} trained on spectra and masses of APOKASC1 stars \citep{pinsonneault_etal_apokasc1_2014}, and derived log age with a precision of 0.2 dex. Adapting our noise model from Eq. \ref{eq:noise_model} with $\sigma_{\mathrm{log}\tau} = 0.2$, and optimizing the likelihood, we find a value for the parameter $x$ of 0.62, which is roughly consistent with our results with the Ting age set: the star formation time-scale decreases slowly towards the outer disk and flattens at $\Ro = 12$ kpc. Additionally, the best fit parameters of the other model aspects are consistent with the work in \cite{frankel_etal_2018}, who used this data set. In particular, the value of the radial orbit migration term was exactly the same: a mixing scale of 3.6 kpc after 8 Gyr. This confirms the robustness of this result, and suggests that our new model aspects and APOGEE selection function are incorporated correctly.

\paragraph{Ness18} This data set also used \textit{The Cannon}, trained on APOKASC2 spectra and masses. We apply the same noise model as for the Ness16 data set. The best fit inside-out parameter for this data set is $x=0.50$, which means that the star formation rate depends more weakly on Galactocentric radius, and goes as $1 - 0.5 \Ro /8\mathrm{~ kpc}$. According to this fit, the star formation rate is as slowly decreasing function of time throughout the disk until $\Ro = 16$ kpc. 

\paragraph{Ness19} \cite{ness_etal_2019} used empirical tight age-abundance relations, and trained \textit{The Cannon} to map from precise abundances measurements for 17 elements to APOKASK2 age estimates. This procedure was applied to the low-$\alpha$ red clump sample of \cite{ting_hawkins_rix_2018}, and yields ages with a precision $\sigma_\tau = 1.6$ Gyr. The resulting distribution of observed ages resembles most that of the Ting18 sample compared to the other age sets.  We modified our noise model accordingly with $p(\tobs ~|~ \tau, \sigma_\tau) \sim \mathcal{N}(\tau, \sigma_\tau)$. The best fit parameter for the inside-out growth on this data set is $x = 1.1$, which implies that the star formation history at 8 kpc has been constant over the past 7.5 Gyr.

\paragraph{Sanders18} We also used the ages presented in \cite{sanders_das_2018}, using the method outlined in \cite{das_sanders_2019}. This method consists of an initial data-driven stage to measure stellar mass from APOGEE spectroscopy as in \cite{ting_rix_2018} and \cite{ness_etal_2016}, but using the information contained in the abundances directly rather than the full spectrum. A second stage compares the derived mass, spectroscopic parameters, photometry and Gaia parallaxes to a set of isochrones folded with a Galactic prior. We first re-run the age determination procedure, removing the priors on the ages. The fit to the data leads to $x = -0.3 < 0$, favouring weak outside-in growth. A key difference between this age determination method and the previous ones is that the luminosity and parallaxes of the stars were explicitly used to derive the stellar masses. This could introduce a bias with distance or extinction, in particular at large distances where the distance information from the parallax is modest and the extinction may be significant. When comparing these ages with all of the previous ones (Ting18, Ness16, Ness18, Ness19), which are not distance-dependent, we find systematic differences that are function of distance: compared to the other data sets, these ages tend to be overestimated at large distances leading to a contaminating population of old stars at large distances. In our case, due to selection effects, distant stars are mainly located in the outer disk. This produced a general trend with older stars at larger distances and larger Galactocentric radii, weakening (and inverting, in some parts) the already quite weak radial age gradient that contains information about inside-out growth in the data set (see Fig.~\ref{fig:age_profile}).

\bibliography{lit}

\end{document}